\documentclass[aps,physrev,amsmath,amssymb, preprint,longbibliography, groupedaddress]{revtex4-2}

\usepackage{graphicx}
\usepackage{bm}
\usepackage{amssymb,amsmath}
\usepackage[version=3]{mhchem}
\usepackage{lmodern,babel,adjustbox,booktabs,multirow}
\usepackage{setspace}
\usepackage{natbib}

\begin{document}

\title{How superradiance survives molecular noise inside virus capsids}

\author{Tianran Li}
\author{Irina B. Tsvetkova}
\thanks{These authors contributed equally}
\author{Matthew Brauser}
\author{Amala Raj}
\author{Grace L. Clemmer}
\author{Philip G. Shushkov}
\email{phgshush@iu.edu}
\author{Bogdan Dragnea}
\email{dragnea@iu.edu}
\affiliation{Department of Chemistry, Indiana University,
Bloomington, IN, 47405, United States}

\date{\today}

\begin{abstract}
Molecular motion generally disrupts the coherence required for cooperative light emission \cite{Gross1982a}. Yet organic dyes attached to icosahedral virus capsids can emit superradiant bursts at room temperature in water\cite{Tsvetkova2019,Holmes2022}. How this collective behavior survives molecular noise remains unresolved. Here we show experimentally that superradiance-active capsids support underdamped, low-frequency vibrations coupled to the dye excited electronic states. Open-system modeling identifies a feedback mechanism in which electronic excitation drives these vibrations through exciton–phonon coupling, while the resulting motion modulates electronic coupling between chromophores. In the model, this modulation dynamically separates bright and dark collective states, suppressing their mixing and restoring superradiant emission despite dephasing. Neither static interchromophore coupling nor a common vibrational modulation of transition energies alone restores cooperative emission under the same conditions. These findings suggest that the capsid acts not merely as a scaffold that arranges emitters, but as a dynamical component whose mechanical motion cooperates with electronic interactions to protect collective emission. This interplay suggests a design principle for molecular architectures that sustain cooperative optical behavior under ambient conditions.
\end{abstract}


\maketitle

\clearpage
\section*{Popular summary}
Quantum effects are notoriously fragile: interactions with the surrounding environment typically destroy the delicate synchronization needed for particles to behave collectively. Yet, the virus particles chemically decorated with fluorescent molecules can produce intense, synchronized bursts of light, a phenomenon known as superradiance, even at room temperature in water. How such collective quantum behavior survives in this noisy environment has remained a mystery. Here, we uncover a surprising role for the mechanical vibrations of the virus itself. Using ultrafast optical measurements, we detect collective vibrations of the virus's protein shell that persist long enough to influence the light-emitting molecules attached to it. Theoretical modeling reveals how these vibrations can dynamically adjust interactions among the molecules, promoting the synchronization required for superradiance despite environmental disturbances. Rather than merely providing a rigid scaffold, the virus shell may actively help sustain collective light emission through its own motion. Our findings suggest an unexpected strategy for designing quantum materials that operate under everyday conditions: instead of eliminating molecular motion, carefully organized vibrations could be harnessed to support collective quantum phenomena.

\section{Introduction}
Quantum materials are transforming information technologies and hold promise for applications ranging from non-invasive imaging and chemical reaction control to sensing and energy harvesting \cite{Arakawa2020, Kavokin2022, Padgett2017, Mandal2023, GardinerZoller2015, Yuan2023, Scholes2010, Tiwari2013}. However, the broader use of quantum behavior is limited by decoherence \cite{Suter2016}; interactions with complex environments---including thermal collisions---rapidly destroy access to macroscopic quantum properties, which is why cryogenic conditions are often required. Nevertheless, because decoherence is governed by the environment, the transition between quantum and classical behavior may itself be tunable. Learning to control this transition is important both for practical quantum technologies and for fundamental questions surrounding measurement and the emergence of classical physics \cite{Zurek2003}. Identifying ways to engineer quantum coherence under ambient conditions could bring quantum technologies closer to everyday use.

Strategies for preserving coherence in molecular quantum systems typically rely on cryogenic temperatures, high vacuum, strong external fields, and precise structural control at microscopic scales. Biological systems operate under very different conditions; yet, several biological processes---including geomagnetic sensing, enzymatic catalysis, and photosynthetic energy transfer---have been proposed to exploit quantum effects under ambient conditions \cite{Ball2011,Kim2021}. These examples raise the intriguing possibility that biological architectures may offer design principles for sustaining quantum behavior in noisy molecular environments \cite{Nagy2006,Kim2021,Babcock2024}. 

\begin{figure*}[!ht]
\centering
\includegraphics[width=\textwidth]{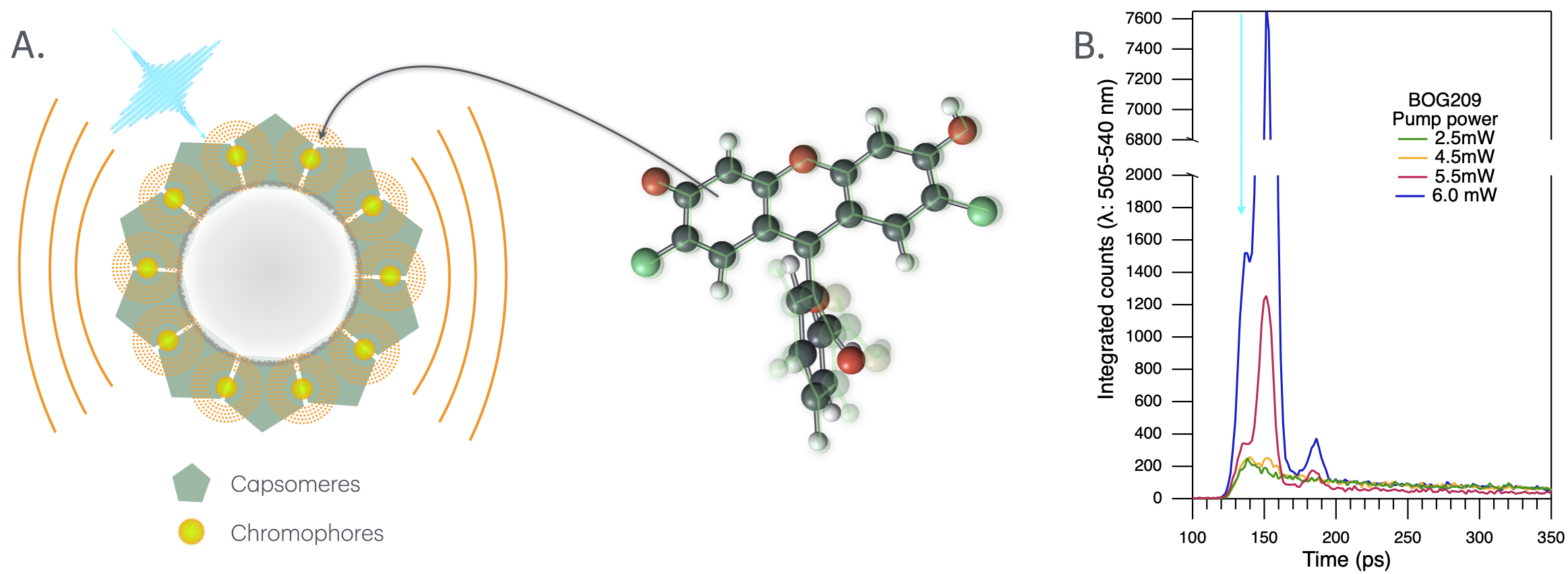}
\caption{\textbf{Virus-bound dyes and superradiant emission.} A. Schematic of a symmetric virus with chromophores wedged in between the oligomeric building blocks (capsomeres) of the virus coat (capsid), and element-colored ground-state geometry (solid colors) of the OregonGreen488 chromophore superposed with its excited-state configuration (translucent) illustrating the structural deformation at excitation. B. Streak-camera measured emission intensity vs time from a BOG sample with $<N>=209$ dyes/particle. The cyan arrow marks the excitation laser pulse. Low excitation power leads to conventional exponential decay; excitation powers higher than a threshold lead to a delayed, short superradiant pulse.}
\label{SRVirusIntro}
\end{figure*}

A particularly revealing test of this possibility is collective quantum-optical emission, which requires coherence among many emitters. In previous work, we observed room-temperature superradiant emission (SR) after ultrafast optical excitation from brome mosaic virus (BMV) capsids densely labeled with Oregon Green dyes, in aqueous solution ~\cite{Tsvetkova2019,Holmes2022} (Fig.~\ref{SRVirusIntro}).  SR is a quantum coherent phenomenon in which indistinguishable emitters coupled to a common electromagnetic field radiate cooperatively as a single macroscopic dipole \cite{Dicke1954,Benedict1996} (Extended Data Fig.~\ref{cartSR}). 

Although a transient phenomenon, SR shares with superconductivity, superfluidity, and polariton condensation the emergence of coherent behavior across many constituents. However, its persistence under ambient aqueous conditions is surprising because environmental fluctuations are expected to destroy the coherence required for collective emission, even on the shortened timescale of the superradiant burst. How can SR survive in such a noisy molecular environment? Here, we combine ultrafast spectroscopy and theoretical modeling to investigate how this cooperative emission survives environmental dephasing. Our results support a feedback mechanism in which electronic excitation drives collective capsid vibrations that, in turn, modulate interchromophore coupling, dynamically protecting the superradiant emission pathway.

\section{Capsid chemistry and architecture}

BMV possesses several chemical and geometric features favorable for SR, including high symmetry, dense chromophore loading, nanoscale confinement, and structural stiffness. The capsid can accommodate up to $\sim$300 Oregon Green 488 (OG) dyes per particle without detectable structural distortion, creating a large ensemble of coupled emitters \cite{Li2026a}. We refer to BMV capsids densely labeled with Oregon Green as BOG particles. 

Reversible capsid swelling \cite{Incardona1964} permits incorporation of chromophores deep within the protein shell, where they experience confined and constrained molecular environments \cite{AnilSushma2021,krishna2025computational}. Capsid icosahedral symmetry promotes chromophore indistinguishability: when the radiation field cannot tell which dye emitted, their emission amplitudes superpose coherently, turning many dyes into one bright collective emitter. This is consistent with the sharp decrease in SR below saturation labeling, where stochastic occupation of the labeling sites breaks the symmetry of the chromophore ensemble. However, these architectural features of BOG particles do not fully explain the room-temperature SR. Specifically, although the swollen capsid preserves global icosahedral symmetry, SR disappears upon the compact-to-swollen transition. Likewise, simple scaling arguments indicate that the observed chromophore number is far below that required for cooperative emission to overcome environmental dephasing through emitter number alone (Appendix Fig.~\ref{cartSR}C). Thus, symmetry, dense labeling, and confinement appear to be necessary but insufficient. This points to an additional ingredient beyond the static structure.

 Thermal vibrational motion generally destroys the phase relationships required for cooperative emission \cite{Gross1982a}. In molecular systems, coupling to phonons leads to dephasing and limits the lifetime of coherence of electronically excited states \cite{Mukamel1995}. However, radiative cooperativity can survive at room temperature, even in submicroscopic and soft matter systems, which are especially vulnerable to disorder, localization, and phonon-driven dephasing. Previous examples include photosynthetic light-harvesting complexes \cite{Monshouwer1997}, nitrogen-vacancy nanodiamonds \cite{Bradac2017}, molecular J-aggregates \cite{Barotov2023}, protein-based systems \cite{Babcock2024}, and hybrid perovskite films \cite{Biliroglu2022, Biliroglu2025}. In several of these systems, vibrational or polaronic effects have been proposed to modulate excitonic delocalization, protect bright states, and/or reduce sensitivity to dephasing \cite{Lim2015,Hestand2018, Biliroglu2025}. Thus, structured environmental fluctuations can actually facilitate quantum behavior by dynamically modulating electronic couplings and energy landscapes \cite{Ishizaki2009,Chin2012,Hestand2018}. Whether vibrational motion can actively promote SR in a complex biological architecture remains largely unexplored.

Here, we show that low-frequency collective motions of dye-labeled BMV capsids can dynamically promote SR. Combining ultrafast spectroscopy and theoretical modeling, we find that phonon-assisted fluctuations transiently create configurations favorable for cooperative radiative decay.

\section{Global capsid motions}

We hypothesized that room-temperature SR is controlled not only by the static geometry of the capsid, but also by how the dye excitations couple to capsid motion. The relevant quantity is the exciton--phonon spectral density of the dye--capsid complex, set by the mode frequencies, damping rates, and coupling strengths. Coherent low-frequency motions of the compact BMV shell could transiently tune chromophore separations, orientations, and site energies into configurations that favor cooperative radiative decay. In other words, coherent phonons could funnel energy into delocalized, bright, SR states, whereas thermal phonons would dephase them. A necessary prerequisite is therefore the presence of organized motion versus thermal bath noise, with underdamped, symmetry-selected capsid modes acting as spectrally selective coordinates that promote SR.

This vibronically-coupled SR requires a directly-detectable feature: the BOGs should exhibit underdamped collective modes that are coupled to the dye excited state on the timescale of cooperative emission.

By virtue of their dense protein and nucleic acid packing, many viral particles are expected to support collective acoustic-like vibrations \cite{vanVlijmen2005, Tsen2006, Hsieh2023, Zhang2025, Kuang2025}. Coarse-grained capsid models suggest that Caspar--Klug quasi-icosahedral capsids \cite{Caspar1962} are distinguished by a vibrational spectrum containing a cluster of 24 low-frequency modes that carry universal group-theoretical content and are separated from the rest of the vibrational spectrum \cite{Englert2008, Peeters2009}. For BMV, this isolated soft-mode manifold provides a plausible route for symmetry-assisted vibronic coupling: simultaneous photoexcitation of multiple capsid-confined dyes changes the local excited-state forces on the dye--protein complex, producing a structural distortion that preferentially projects onto the soft collective coordinates. This hypothesis is conceptually related to the quantum analog of vibration isolation (QAVI) invoked to explain room-temperature superfluorescence in hybrid perovskite films, where polaron formation was proposed to protect collective electronic polarization from thermally driven dephasing \cite{Biliroglu2022}. In BOGs, the analogous polaron-like protective element involves the symmetry-isolated soft-mode manifold of the icosahedral capsid.

It is important to make the distinction between underdamped and overdamped capsid modes. An underdamped mode provides a spectrally narrow, phase-persistent collective coordinate that can coherently modulate the dye Hamiltonian during the SR build-up time. An overdamped mode, in contrast, would broaden dye transition energies, randomize excitonic phases, and convert symmetry-organized vibronic coupling into incoherent dephasing. Therefore, measuring the frequencies and damping of electronically coupled capsid modes provides a direct test of whether capsid phonons can participate in RT-SR.

The frequencies, damping rates, and electronic-state coupling of the symmetry-isolated soft modes of an icosahedral dye-labeled capsid have not been experimentally determined to the best of our knowledge. Although recent single-particle optical measurements on larger, non-icosahedral virus particles have been interpreted as evidence of acoustic-like vibrations \cite{Zhang2025}, they did not address the symmetry-specific soft-mode manifold predicted for Caspar--Klug icosahedral capsids. 

In BOGs, ultrafast excitation can launch capsid motion because electronic excitation changes the local forces within the dye--protein complex, Fig.~\ref{SRVirusIntro}B, hypothetically producing a displacive impulse along capsid normal coordinates, similar to the excitation of coherent phonons in crystalline solids \cite{Zeiger1992, Dekorsy2000}. Photon recoil and hydrodynamic anisotropy may provide additional weak impulsive forces. In the modeling section, we considered that the dominant coupling arises from excited-state structural changes (Appendix ~\ref{SI:Identification}).

\subsection{Optical detection of acoustic modes}

\begin{figure*}[t]
\centering
\includegraphics[width=\textwidth]{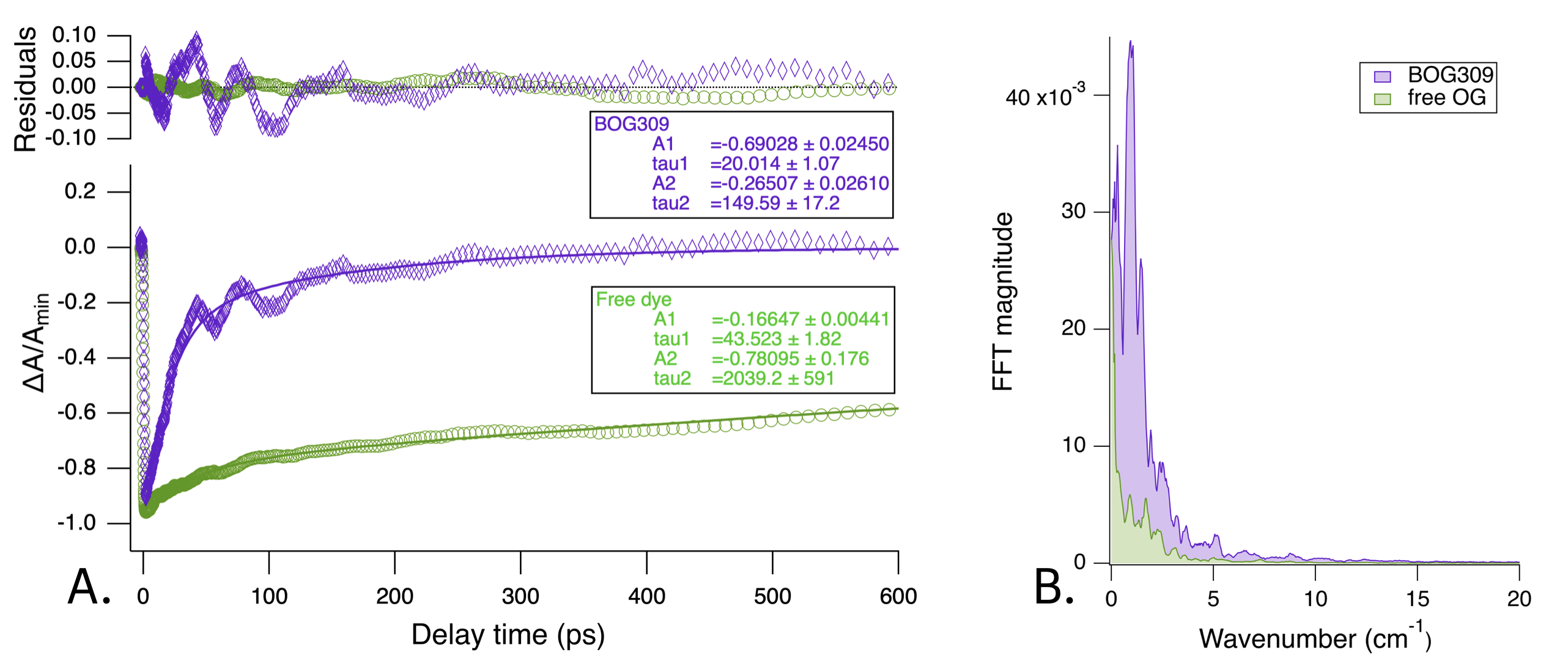}
\caption{\textbf{Transient absorption of free and virus-bound dyes.} A. Transient absorption traces for a 70 nM solution of BOG particles carrying $N \approx 309$ dyes per particle (purple diamonds), and a free OG solution (green circles) of the same optical density, and bi-exponential fit curves (parameters in the inset), showing evidence of TA oscillatory modulation in the residuals. B. Fast Fourier transforms of the raw TA data for BOG309 and the free OG control.}
\label{tabogog}
\end{figure*}

To test whether optical excitation launches underdamped global vibrational modes, we performed transient absorption (TA) spectroscopy on solutions of BOG particles, Fig.~\ref{tabogog} (Methods and Appendix Figs.~\ref{setup}--\ref{S:BOG309TA}). In the TA measurements, the ground-state bleach reports the recovery of the dye ground-state population, and therefore the decay of the optically prepared excited ensemble. For BOGs, bleach recovery is dominated by two exponential components. The fast component, with a relaxation time of a few tens of picoseconds, is assigned to SR damping (Fig.~\ref{SRVirusIntro}B) and solvent relaxation. The slower component, with a relaxation time ranging from hundreds of picoseconds to several nanoseconds, is assigned to the conventional, uncorrelated population relaxation from dyes that do not participate in the SR burst.

At the 70 nM total dye concentration used in Fig.~\ref{tabogog}, each excitation pulse interacts with \(\sim10^7\) BOGs within the pump--probe overlap volume. Randomly phased particle shape fluctuations effectively average out. Any oscillatory signal that survives ensemble averaging must instead arise from a vibrational response launched with a reproducible phase relative to the pump pulse. After the biexponential population recovery from the TA traces is subtracted, the residuals reveal a small but reproducible oscillatory modulation of the absorbance. These oscillations are already visible in the raw BOG TA traces, whereas they are absent in the free OG control and in samples with low dye-to-particle ratios, both in the raw data and in the residuals obtained after subtraction of the same type of fit function. A Fourier component near \(0.9\)--\(1.0~\mathrm{cm^{-1}}\) was reproducibly detected in independent dye-saturated BOG samples (see Appendix Fig.~\ref{S:BOG309TA}). This frequency lies in the range expected for low-frequency acoustic-like modes of a small hydrated virus capsid \cite{Dykeman2010, Burkhartsmeyer2020}.

\section{Core--shell mechanical transduction}

The TA measurements reveal underdamped oscillations at early delay times in dye-saturated BOG samples, but not in free OG controls or in weakly labeled particles. To determine whether this oscillatory response reflects an intrinsic mechanical property of the BMV scaffold, rather than a feature introduced solely by the covalently attached dyes or the RNA cargo, we next examined BMV protein cages assembled without OG chromophores.

\begin{figure*}[t]
\centering
\includegraphics[width=\textwidth]{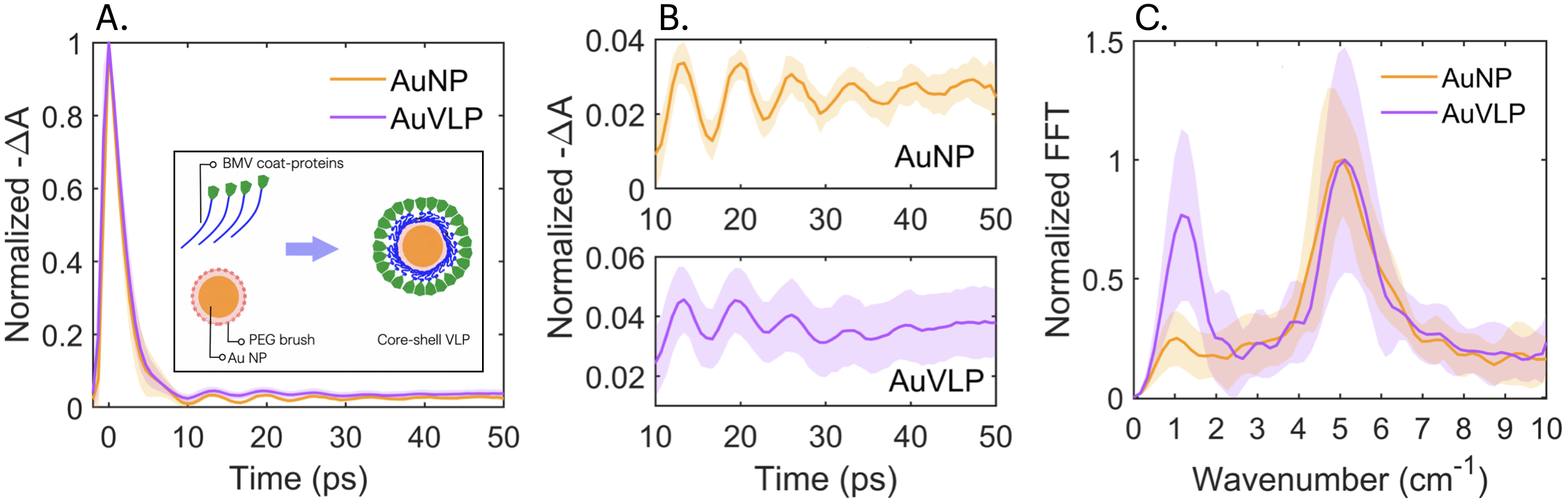}
\caption{\textbf{Probing capsid vibration via AuNP breathing mode.}
		A.  Averaged transient-absorption traces for AuNP and AuVLP; inset: cartoon of core-shell VLP assembly from BMV coat proteins and PEG-coated NPs. The positively charged N-termini (blue) of the BMV coat protein bind to the negatively charged PEG--coated NP surface. Hydrophobic and electrostatic interactions lead to an icosahedral tiling of the sphere with coat proteins. B. Zoomed-in view of A highlighting acoustic oscillations. C. Normalized, averaged Fourier transforms of the detrended TA traces for AuNP and AuVLP. In all graphs, shaded areas represent standard deviations from multiple experiments.}
\label{CoreShell}
\end{figure*}

Direct optical detection of acoustic-like modes from empty protein cages in solution is challenging because the protein shell has a weak optical response. We therefore used a gold nanoparticle as an opto-mechanical transducer. Unmodified BMV coat proteins were assembled around spherical Au nanoparticles coated with a short polyanionic brush, forming icosahedral core--shell virus-like particles (AuVLPs), Fig.~\ref{CoreShell}A (Methods and Appendix Fig.~\ref{auvlp}) \cite{Sun2007}. 

In this geometry, the Au core provides a strong TA signal, while the surrounding protein shell can modify the frequency and damping of the acoustic response associated with the core through mechanical coupling. Comparing the TA response of the bare Au NP cores with that of AuVLPs therefore provides an independent way to detect underdamped capsid-associated modes, analogous to previous studies of mechanical coupling between metal nanoparticle cores and surrounding dielectric shells \cite{Fernandes2018, Wang2023}.

This transduction strategy relies on the well-established ultrafast acoustic response of Au NPs. After pump absorption, the conduction electrons in the Au core rapidly thermalize through electron--electron scattering and then transfer energy to the lattice through electron--phonon coupling on a timescale of \(\sim 1\) ps \cite{Link1999,Hodak2000}. The resulting impulsive thermoelastic stress launches coherent radial expansion and contraction of the Au lattice, corresponding to the \(l=0\) breathing-mode acoustic phonon \cite{Hartland2006}. When the BMV capsid is tightly assembled around the Au core, the core breathing motion applies a periodic radial stress to the surrounding protein shell. The AuNP therefore acts as an optoacoustic transducer: its strong, impulsively launched breathing mode can mechanically drive, hybridize with, and be damped by capsid-associated acoustic modes. Consequently, changes in the frequency, phase, or damping of the AuNP breathing oscillation upon capsid assembly provide an indirect but sensitive probe of the vibrational modes of the BMV shell.

Transient absorption measurements of bare AuNP cores (20 nm diameter) showed the expected prompt increase in transmission associated with bleaching of the surface plasmon resonance, followed by a clearly resolved coherent oscillation (Fig.~\ref{CoreShell}A). The oscillation period was \(\sim 6.5~\mathrm{ps}\), corresponding to \(5.1~\mathrm{cm^{-1}}\), in agreement with previous measurements and with the predicted \(l=0\) breathing-mode frequency for AuNPs of this size \cite{Hartland2006}. Thus, the bare AuNP provides an internal calibration for the optoacoustic transduction experiment.

When the same AuNP cores were encapsulated in a BMV capsid, the \(5.1~\mathrm{cm^{-1}}\) Au breathing mode remained visible, indicating that the core still undergoes coherent radial expansion and contraction. However, the AuVLP traces exhibited an additional, slower modulation superimposed on the AuNP breathing oscillation (Fig.~\ref{CoreShell}B). Fourier analysis of the detrended 10--50 ps time window revealed two spectral components in the AuVLP samples: the intrinsic AuNP breathing mode at \(\approx 5.1~\mathrm{cm^{-1}}\), and an additional low-frequency peak at \(\approx 1.3~\mathrm{cm^{-1}}\) (Fig.~\ref{CoreShell}C). The appearance of this low-frequency component only after capsid assembly shows that the signal is not an intrinsic AuNP breathing harmonic, or an artifact from chromophore conjugation, but a mechanical response introduced by the protein shell (Methods, AuVLP transient-absorption controls).

To derive an estimate for the frequency of the capsid mode and the capsid--NP coupling strength from the data, we modeled the Au core and capsid as two coupled damped harmonic oscillators (Appendix ~\ref{SI:CoupledOscillators}). This minimal two-mass, two-spring model yields an effective core--capsid coupling constant $\approx 200$-fold smaller than the AuNP breathing-mode spring constant, together with a damping coefficient of the capsid mode (Appendix ~\ref{SI:CoupledOscillators} and Appendix Table~\ref{edt:core}b). Thus, the AuVLP experiment provides an independent mechanical readout of an underdamped, low-frequency capsid mode in the same spectral range as the oscillatory residuals observed in dye-saturated BOGs.

The present measurements do not uniquely assign the oscillation to a specific normal mode, but the observed frequencies fall in the range expected for the lowest spheroidal acoustic-like modes of a hydrated BMV capsid (Appendix ~\ref{SI:Identification}). Accordingly, we assign the observed oscillations to capsid-associated acoustic-like modes rather than to a specific member of the 24-mode manifold. Moreover, for dye-saturated BOGs, the observed frequency at ($\sim 1~\mathrm{cm}^{-1}$) is close to the reported $1.07~\mathrm{cm}^{-1}$ for a single 25~nm icosahedral virus in an experimental study using extraordinary acoustic Raman spectroscopy~\cite{Burkhartsmeyer2020}.

\section{Vibrationally assisted superradiance}

The experiments above establish two facts that are difficult to reconcile with a purely static picture: low-frequency capsid modes are underdamped in BOGs and AuVLPs, and, in BOGs, capsid motion is coupled to the OG excited state. We therefore asked whether a capsid vibration could do more than perturb the dyes. Could it help a noisy dye ensemble recover the collective polarization required for SR?

To address this question and reveal the possible physical mechanism, we used a minimal open-system model of a dye shell coupled to one collective capsid coordinate (Appendix ~\ref{SI:Qmodel}). Each chromophore was treated as a two-level emitter, and SR was represented by a collective emission channel in which phase-correlated dyes radiate through the same optical mode. Environmental noise was included as uncorrelated fluctuations of the individual dye transition energies, representing local overdamped disorder from the hydrated capsid environment. This dephasing term destroys SR in the model, as expected.

The collective, underdamped capsid motion was represented by a single coordinate, \(x(t)\), with a frequency and lifetime chosen from the experimentally observed and calculated low-frequency capsid-mode range (Appendix ~\ref{SI:Qmodel}). We allowed this coordinate to affect the dye shell in two distinct ways. First, the vibration could modulate the transition energies of individual dyes, corresponding to direct exciton--vibrational coupling. Second, the vibration could modulate the coupling between dyes, because capsid motion changes chromophore separations, orientations, and therefore dipole--dipole interactions. This second pathway represents a vibrationally driven modulation of the excitonic interaction network.

\begin{figure*}[!ht]
\centering
\includegraphics[width=0.77\textwidth]{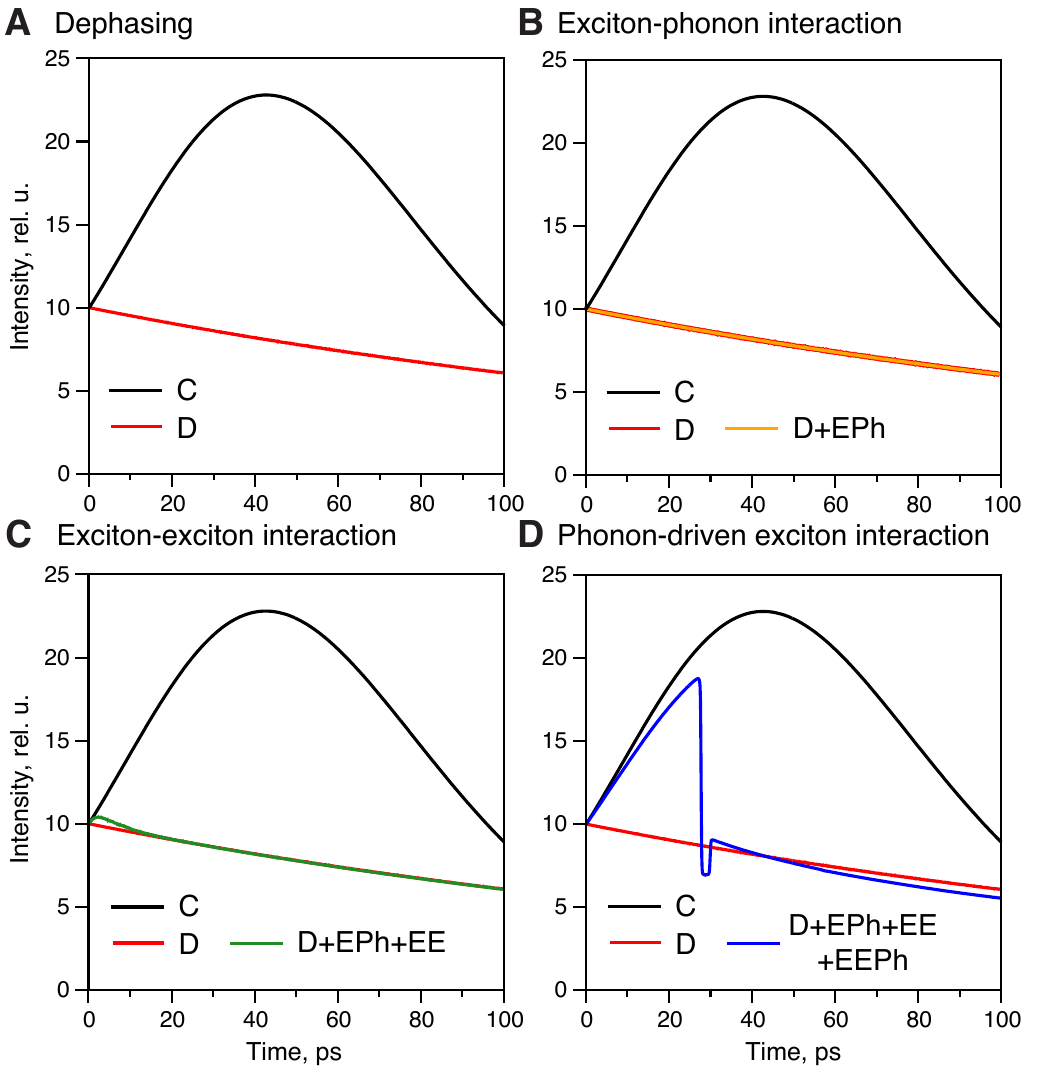}
\caption{\textbf{Vibronically assisted recovery of superradiance.}
	Simulated emission from a 10-chromophore model. A. Black: fully coherent evolution. Red: dephased ensemble. B. Yellow: dephasing plus direct exciton--phonon coupling. C. Green: dephasing plus direct exciton--phonon coupling and static exciton--exciton coupling. D. Blue: dephasing plus phonon-driven modulation of exciton--exciton coupling. Only the final case restores an SR-like burst.}
\label{mechanism_superradiance}
\end{figure*}

For computational tractability, the simulations were performed for a 10-chromophore system. The purpose was not to reproduce the full number of dyes on a BOG particle, but to identify which physical coupling pathway is capable of restoring cooperative emission in the presence of molecular noise. In the absence of dephasing, the model produces the expected delayed SR peak, Fig.~\ref{mechanism_superradiance}A. When uncorrelated dephasing is included, this peak disappears and the emission becomes incoherent. Adding a common underdamped vibration that only shifts dye transition energies does not recover SR, Fig.~\ref{mechanism_superradiance}B. Including static exciton--exciton interactions is also insufficient, Fig.~\ref{mechanism_superradiance}C. SR-like emission reappears only when the underdamped vibration modulates the exciton--exciton interaction itself, Fig.~\ref{mechanism_superradiance}D.

These results suggest that the capsid mode dynamically reshapes the electronic coupling network of the dye shell. Recent work suggests that the reshaping will be rapidly followed by a redistribution of the excitation energy by FRET \cite{Li2026a}. Efficient transfer of energy between emitters was noted as a favorable factor for cooperative radiation processes in a recent example of room-temperature super-fluorescence from up-converting nanocrystals\cite{HuangK2022}.

During parts of the vibrational cycle, the chromophore ensemble is driven toward configurations with larger collective oscillator strengths, allowing a burst of cooperative emission to emerge despite ongoing dephasing. The calculated emission displays a delayed, shortened SR burst and an oscillatory interference pattern at its trail end set by the coherent capsid motion (see Fig.~\ref{mechanism_superradiance}D) at $\sim 50$ ps. Thus, the model also predicts an oscillatory modulation of the SR-like emission after the initial burst. Indeed, consistent with this prediction, the streak-camera measurements of BOG209 show a secondary emission feature near \(55~\mathrm{ps}\) after excitation only at pump powers that produce SR-like emission, Fig.~\ref{SRVirusIntro}B.

Thus, the experimental data establish the presence of pump-launched, underdamped capsid-associated modes, whereas the model identifies a physically specific route by which such modes can promote SR: dynamic modulation of dye--dye coupling. According to the proposed vibronic mechanism, underdamped capsid phonons can promote RT-SR when they modulate the interactions among dyes, rather than merely shifting individual dye transition energies.

A further reduced, two-chromophore model based on the same ideas makes the mechanism of vibrationally--assisted SR transparent (Appendix~\ref{SI:TwinSync}). In the one exciton manifold, the two emission pathways combine to form a symmetric bright state, $|+\rangle = 1/\sqrt{2}(|10\rangle+|01\rangle)$ and an anti-symmetric dark state, $|-\rangle = 1/\sqrt{2}(|10\rangle-|01\rangle)$. Due to the indistinguishability of the photons emitted by the two chromophores, collective radiative decay from the doubly excited state proceeds through $|+\rangle$. A difference between the chromophore transition energies couples the dark and bright states, disrupting the symmetric superposition and driving the emission towards the independent emitter limit. However, excitonic coupling has the opposite effect: it separates the bright and dark states by an energy $2 \hbar \Omega_{EE}$, which suppresses their mixing, thus protecting the superradiant pathway. A synchronous shift of both transition energies by the acoustic mode moves the two states together without changing their symmetry. Therefore, it does not suppress SR. 

\section{Summary}

In summary, photoexcitation displaces an underdamped collective capsid coordinate, launching a coherent vibration that feeds back on the chromophores by modulating their excitonic coupling and, consequently, the bright--dark energy gap. Near the turning points of the vibrational motion, this gap is largest, dark-state admixture is minimized, and superradiant emission is enhanced; near the equilibrium position, the gap is smallest and the emission weakens. The same vibration collectively modulates the chromophore transition energies, producing the oscillations observed in transient absorption, while its modulation of the bright--dark splitting produces the corresponding oscillations in superradiant emission.

Taken together, the findings are suggesting a picture in which the BMV capsid is not a passive scaffold for arranging dyes, but an active dynamical element in room-temperature superradiance. Compact, dye-saturated capsids support underdamped acoustic-like modes in the \(\sim 1~\mathrm{cm^{-1}}\) range, observed directly in BOG transient absorption and independently through AuNP--capsid mechanical transduction. The modeling shows that such motion can recover an SR-like burst only when it modulates the interactions among dyes. A bi-directional exciton--vibrational feedback dynamically protects collective emission against stochastic emission. Thus, the survival of SR at room temperature in water may not require eliminating molecular noise, but organizing part of it into a symmetry-guided collective coordinate that transiently increases the oscillator strength of the dye ensemble. Icosahedral virus capsids therefore suggest a design principle for ambient quantum materials: biological architectures can use their own soft mechanical modes to convert environmental motion from a source of decoherence into a pathway for cooperative emission.

\section*{Acknowledgements}
The authors gratefully acknowledge Dr. Richard D. Schaller at the Center for Nanoscale Materials -- Argonne National Laboratory, for access to the time-resolved optical spectroscopy instrumentation, Dr. W.L. Schaich for insightful discussions, and Lily Delalande for help with writing a custom code for TEM data analysis of VLP morphology.

The work was supported by grant awards to B.D. from the US National Science Foundation CBET \#: 2232717 and 2232718, and by the Materials Design Program at the Army Research Office, through award \# W911NF2410225. I.T. and B.D. are grateful for partial support from the Romanian Ministry of Research, Innovation, and Digitization under PNRR-I8/C9-CF105 - contract no. 760099.  P.S. and M.B. acknowledge partial support from start-up funds from Indiana university. This research was supported in part by the Lilly Endowment, Inc., through the Indiana University Pervasive Technology Institute.

\section*{Author contributions}
I.T. and B.D. designed the experiments. T.L., I.T. and A.R. performed the experiments. T.L., I.T., A.R. and B.D. analyzed the experimental data. M.B and P.S developed the model for capsid vibrationally-assisted superradiance and performed the simulations. T.L., I.T., M.B., P.S. and B.D wrote the manuscript, with input from all authors.

\section*{Competing interests}
There are no competing interests to declare.

\section*{Code availability}
Code for simulations is available at \url{https://github.com/matt24brauser-cmyk/super_radiance_virus}.

\clearpage
\appendix
\setcounter{figure}{0}
\setcounter{table}{0}
\renewcommand{\figurename}{Supporting Figure.}
\renewcommand{\tablename}{Supporting Table}

\section{Methods}\label{methods}
\subsection{Synthesis of dye-labeled BOGs}
Dye-labelled viruses were prepared according to previously reported protocol \cite{Tsvetkova2019,AnilSushma2021}. Briefly, BMV  was diluted in sodium bicarbonate buffer (100 mM, pH 7.5) to concentration of 10$^{13}$ particles/ml. Oregon Green 488 carboxylic acid, succinimidyl ester was freshly dissolved in DMSO and mixed with BMV solution at ratios ranging from 5000 to 20000 dyes/virus. The reaction was left for 2-3 hours at room temperature, then unreacted dye was removed via filter wash with SAMA buffer. The wash step was done until there was no measurable presence of free dye in supernatant by UV-Vis. Particle concentration and labeling efficiency was measured by UV-Vis. Approximately 200 \textmu{}L of refrigerated BOG solution at concentration of 10$^{13}$ particles/ml was used for TA spectroscopy experiments.

\subsection{Synthesis of AuVLP}
Citrate stabilized AuNP were synthesized through seeded growth \cite{Liu2012GNP}. Citrate was exchanged to carboxylated TEG ligand and particles were purified and concentrated by ultracentrifugation \cite{Tsvetkova2015}.
AuVLPs were synthesized according to the published protocol \cite{Sun2007, Tsvetkova2015} (Fig.~\ref{CoreShell},inset), with the modifications to account for larger core size. Purified BMV capsid protein dimers (CPDs) were mixed with functionalized AuNP at the ratio of 350 dimers per NP to ensure excess of proteins in the reaction. The final volume of reaction was adjusted to 0.5 mg/ml of proteins. Two step --12 hours each-- dialysis against TNKM (0.05M Tris-HCl, 0.05M NaCl, 0.01M KCl, 0.005M MgCl$_2$, pH 7.4) and then SAMA (0.05M NaOAc, 0.008M Mg(OAc)$_2$, pH 4.6) buffers was done to reassemble BMV capsid around AuNP cores. 

\subsection{Electron microscopy of AuVLPs}
\begin{figure*}[!ht]
\centering
\includegraphics[width=0.9\textwidth,height=170mm,keepaspectratio]{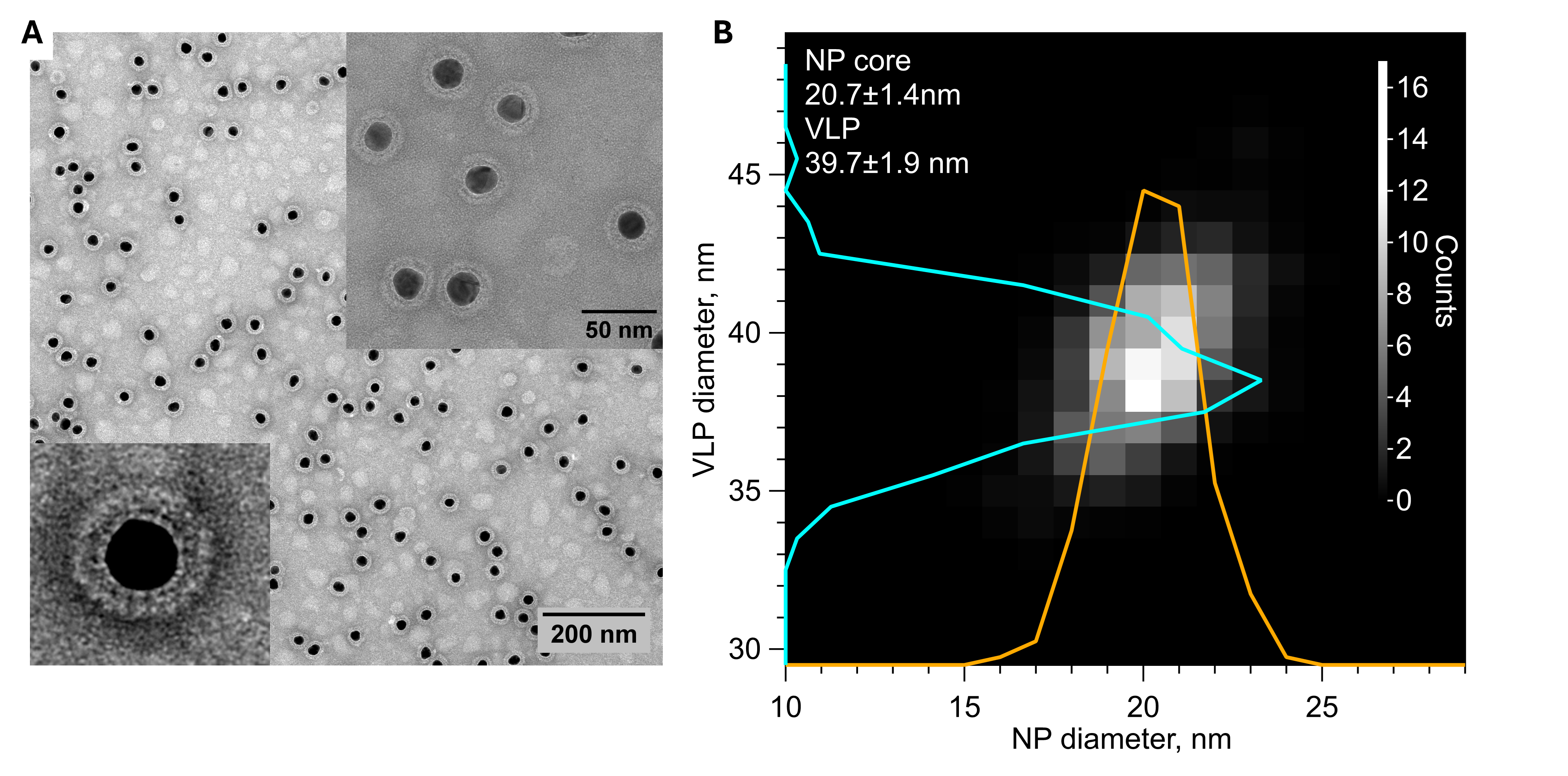}
\caption{Transmission electron microscopy characterization of AuVLPs with 20 nm Au cores. (A) Representative TEM image of AuVLPs. Inset: Protein shell shows evidence for capsomeric organization. (B) Core/VLP diameter double histogram.}
\label{auvlp}
\end{figure*}

Sample preparation for TEM was done by placing 10 \textmu{}L of 1:10 diluted VLP solution on a carbon-coated copper grid. After 5 min, the excess solution on the grid was removed with filter paper and  10 \textmu{}L of 2\% uranyl acetate was used to stain for 1 min. Excess solution was removed by blotting with filter paper. The sample was then left to dry for 10-15 minutes. Images were acquired at an accelerating voltage of 120 kV on a JEOL 1400 transmission electron microscope.
The size of gold nanoparticles core and the diameter of VLPs was measured from the transmission electron microscope (TEM) images of AuVLPs (Appendix Fig.~\ref{auvlp}A). The AuNP core has a diameter of $20.7\pm1.4$~nm, and the overall AuVLP measures $39.7\pm1.9$~nm.

Considering that the BMV coat protein forms T=3 capsids with a 28 nm outer diameter, the closest Caspar-Klug capsid formed of BMV coat proteins consistent with the 39.7 nm diameter and within the TEM error limit would have a T=4 structure. However, note that strained closed shells can form outside canonical Caspar-Klug structures\cite{Bond2020}. Because the Au core imposes a larger core--shell geometry than native BMV, the AuVLP experiment is not intended to reproduce the exact normal-mode spectrum of BOGs. Instead, it tests whether BMV coat-protein shells can support underdamped low-frequency acoustic-like modes that mechanically couple to an optical transducer.

Data were analyzed in ImageJ and Python: core diameters were calculated from area measured from threshold-masked image, then VLP diameter was found from radially averaged profile.

\subsection{AuNP and AuVLP absorbance} 
\begin{figure}[ht]
\centering
\includegraphics[width=0.5\textwidth]{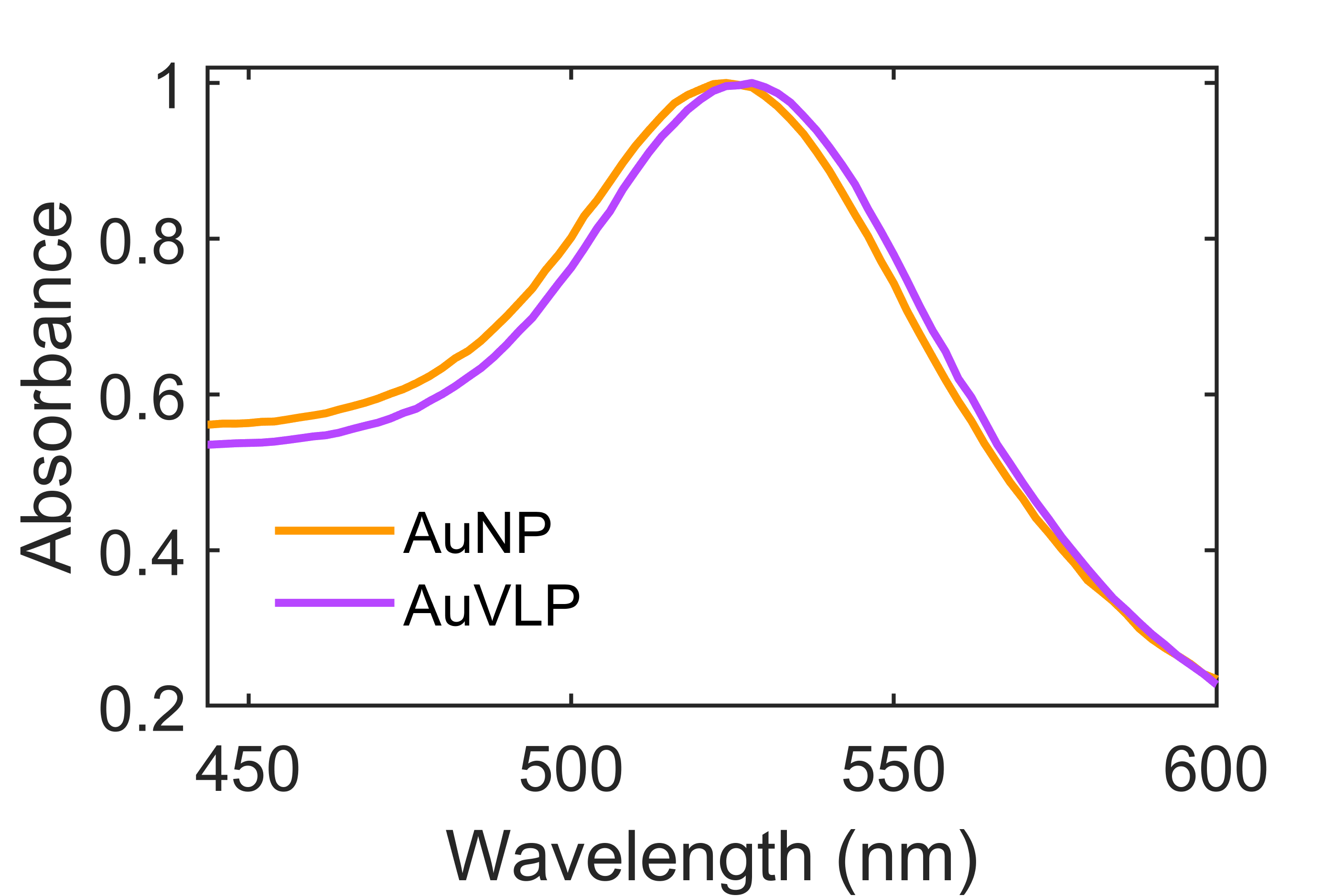}
\caption{Absorption spectra for AuNP/AuVLP. Spectra were normalized to unity at maximum absorbance.}
\label{abs}
\end{figure}

Appendix Fig.~\ref{abs} presents the UV--vis absorption spectra of AuNP and AuVLP. The peak resonance of AuNP at 524 nm shifts to 528 nm in AuVLP, reflecting changes in the dielectric environment. This 4 nm spectral shift is smaller than the 10 nm spectral width of the probe beam in TA experiments. 

\subsection{Samples for optical spectroscopy}
The concentrated AuVLP sample was diluted to approximately 200 \textmu{}L to an optical density at 520nm of 1.0 immediately before the experiment. The dilution solvent was water for AuNPs and SAMA buffer (pH 4.6) for AuVLPs for the stability of viral capsid.

\subsection{Time-resolved emission spectroscopy}
\begin{figure*}[t]
\centering
\includegraphics[width=\textwidth,height=170mm,keepaspectratio]{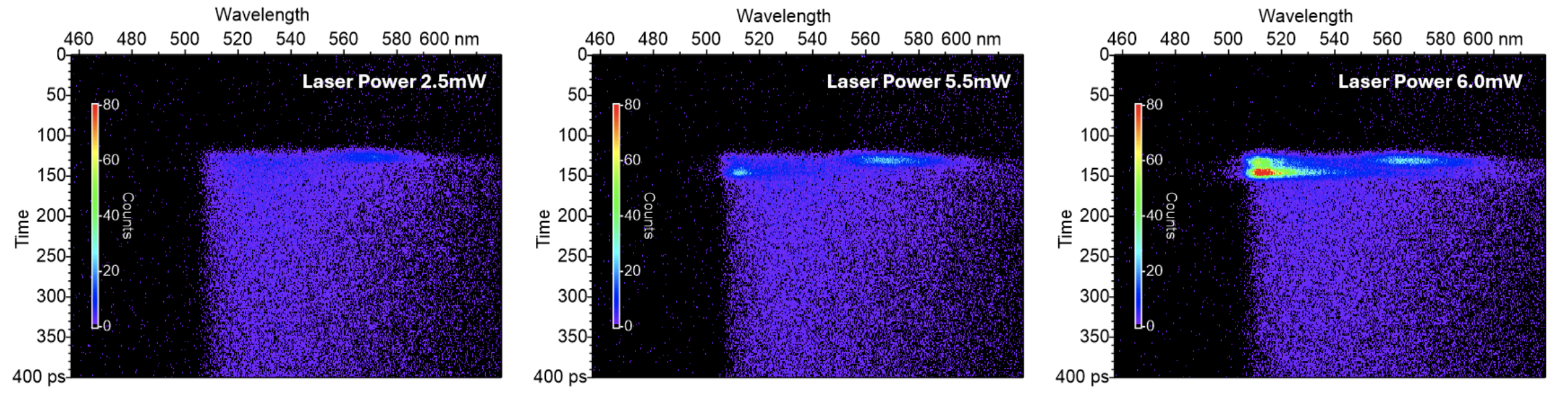}
\caption{Streak camera detected emission from BOG209 at different pump powers.}
\label{streakBOG209SI}
\end{figure*}

Time-resolved fluorescence emission measurements were performed as described previously\cite{Holmes2022}. Specifically, samples in a 1 mm quartz cuvette were excited at 480 nm using $\sim$35 fs pulses generated by an optical parametric amplifier pumped by a Ti femtosecond laser (Spitfire Pro, Spectra-Physics) operating at a repetition rate of 2 kHz. The excitation beam was focused at the sample to a spot with diameter $\sim$160 $\mu$m and the average incident power was varied from 1.0 to 6.5 mW, corresponding to pulse energies of 0.5--3.25 $\mu$J. The beam was directed onto the cuvette at a small angle to minimize the cavity effects. Fluorescence was collected with a lens, passed through an LP508 long-pass filter, dispersed by a 150 mm spectrograph, and recorded  using a photon-counting streak camera (Hamamatsu C5680) as a function of both emission wavelength and time after excitation. Data were preprocessed using HPDTA software, followed by further processing and analysis in Igor Pro (Appendix Fig.~\ref{streakBOG209SI}, Fig.~\ref{SRVirusIntro}B).

\subsection{Transient absorption}
\label{SI:TA_setup}

\begin{figure*}[!ht]
\centering
\includegraphics[width=0.8\textwidth,height=170mm,keepaspectratio]{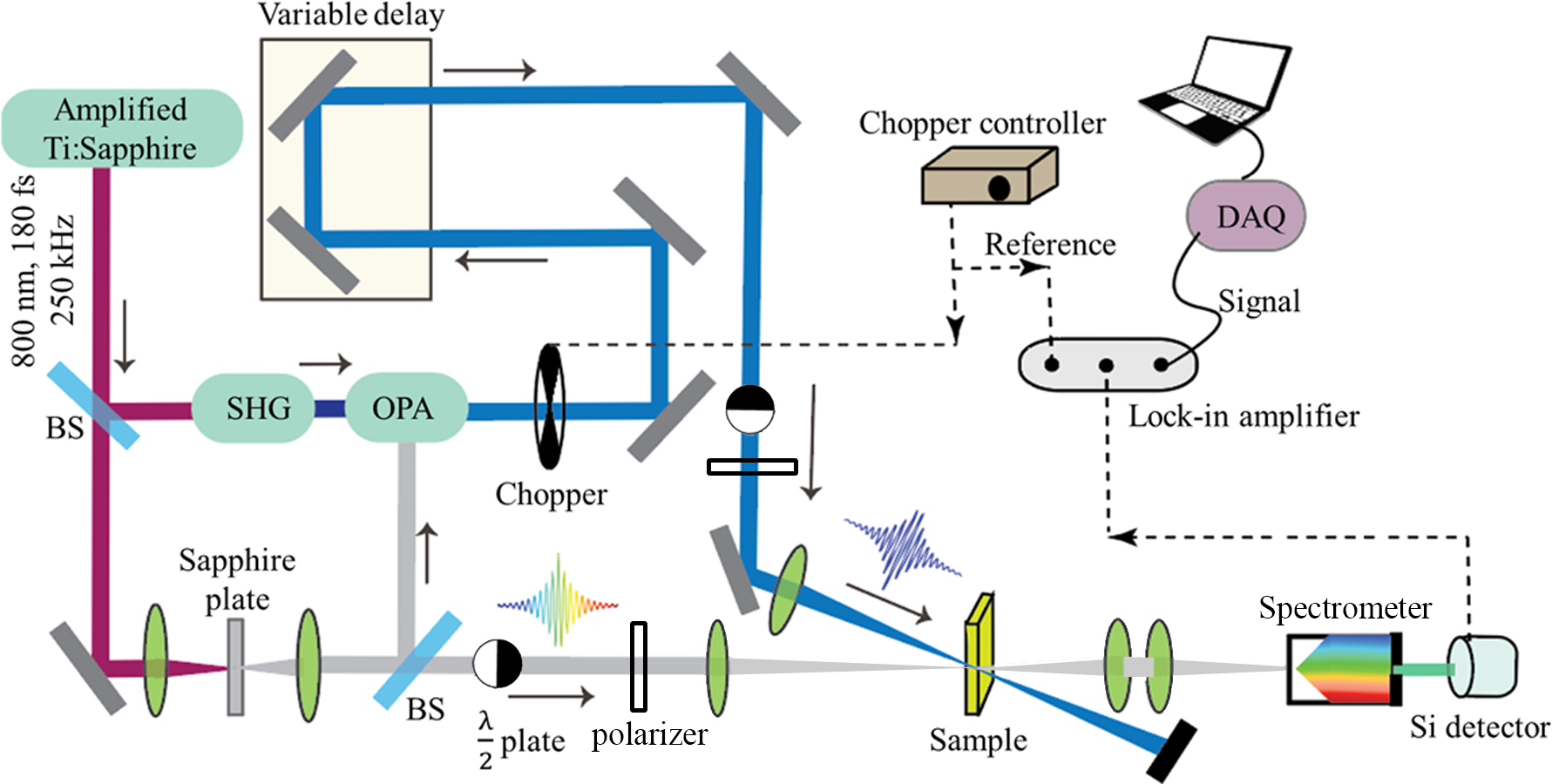}
\caption{\textbf{Schematic of transient absorption setup.}
		Red: 800~nm output from Amplified Ti: Sapphire. Blue: 490~nm pump beam. Gray:  white-light continuum.}
\label{setup}
\end{figure*}

The experimental setup for transient absorption is shown in Appendix Fig.~\ref{setup}. 

The signal and white-light continuum from a double-pass optical parametric amplifier served as the pump and probe beams, respectively. The pump and probe pulses had a duration of \(\approx 170\)~fs at a repetition rate of 250~kHz. The pump pulse energy was \(\sim 10\)~nJ, and its spectrum had a full width at half maximum (FWHM) of \(10\,\mathrm{nm}\). The pump--probe delay was controlled by a computer-driven translation stage (PI M-505) with an accuracy and unidirectional repeatability of 0.25\,\textmu m and a travel range of 150\,\text{mm}. 

\begin{figure}[h]
\centering
\includegraphics[width=0.5\textwidth,height=170mm,keepaspectratio]{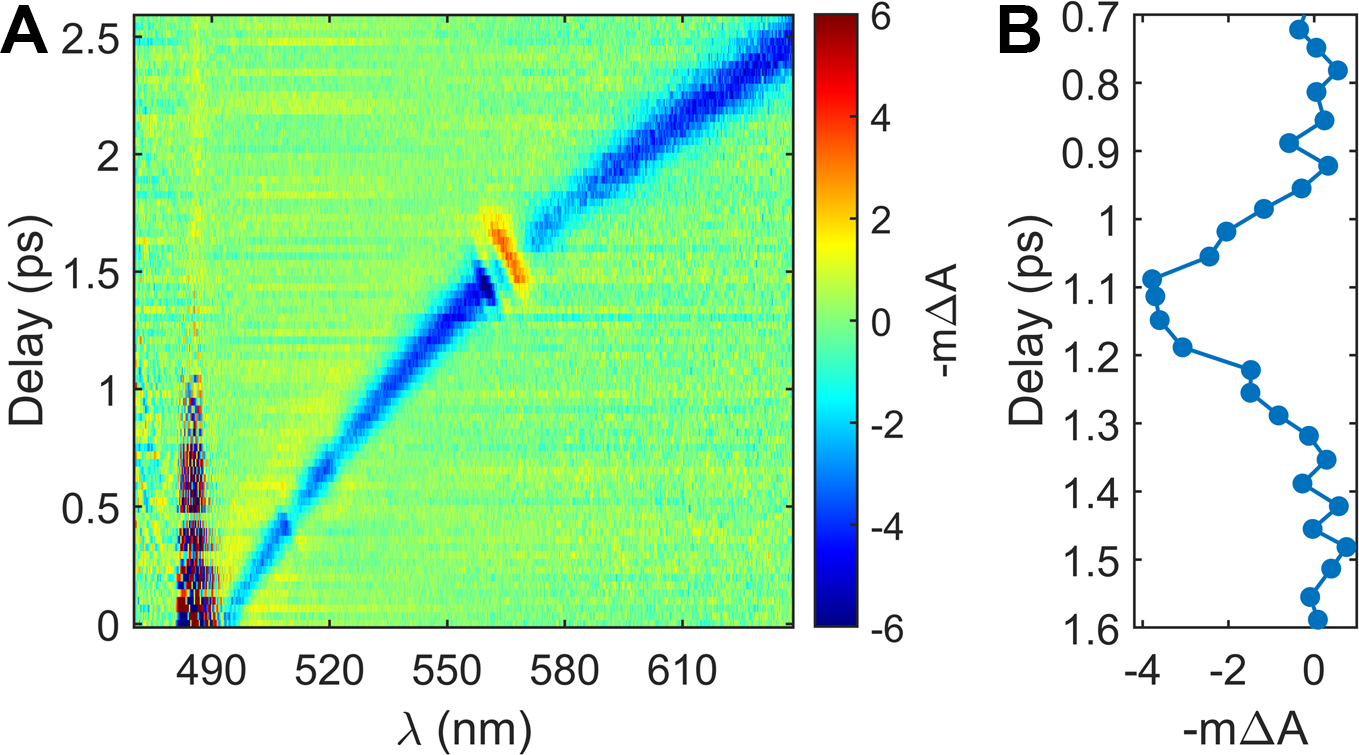}
\caption{\textbf{Instrument response function (IRF) characterization.} A. TA of a broadband probe in methanol. B. Kinetic trace at 545 nm. }
\label{metta}
\end{figure}

The instrument response function (IRF) of the setup, limited by pulse duration and dispersion, was determined via cross-phase modulation (XPM) in methanol~\cite{RUCKEBUSCH20121} (Appendix Fig.~\ref{metta}A). 

The pump wavelength was \(488\,\mathrm{nm}\), and the pump--probe polarizations were set at the magic angle. 
The 545 nm kinetic trace exhibits an XPM feature with a FWHM of \(\approx 200\)~fs (Appendix Fig.~\ref{metta}B), which defines the temporal resolution of the setup\cite{ekvall2000,Lorenc2002}. The signal around 570 nm is a Raman response attributed to the symmetric \(\mathrm{CH_3}\) stretch of the methanol\cite{Tortschanoff,emin2020}. The spike-like feature near the pump wavelength is caused by coherent pump--probe interference\cite{vonJena1979}. 

The sample solution was loaded into a standard quartz cuvette (1 mm path length) and stirred using a custom-made Teflon-coated magnetic wire stirrer. Pump and probe were focused by 50~mm and 70~mm lenses, respectively. The pump-beam focus size was estimated to be \(\approx 110~\mu\mathrm{m}\) diameter from transmission through a set of calibrated diameter pinholes. The pulse peak power was estimated as $0.65$ GW/cm$^2$. For an AuNP at the focus, approximately \(0.47\%\) of the atoms are excited by a single pump pulse. The collection optics following the sample consisted of a pair of 50 mm lenses. The probe was dispersed and wavelength-selected by a SpectraPro-300i spectrometer (Acton Research Corp.) fitted with a 300~lines/mm grating. The pump beam was intensity-modulated at 1~kHz with a mechanical chopper, and the signal was demodulated with an SR830 DSP lock-in amplifier referenced to the chopper; the integration time was 1~s per delay point.

For BOG TA experiments, the pump beam was directed opposite to the probe beam to minimize pump laser scattering in the detection path and maintain consistency with the Argonne streak-camera setup. A $490 \pm 10$ nm bandpass filter with a rejection ratio of $10^6$ was inserted into the probe path in front of the sample. Probe intensity was measured directly after the sample by a detector placed behind the collection lens. Because the dye exhibits much lower molar absorptivity than the AuNPs, a BOG concentration approximately two orders of magnitude higher was required for a similar optical density. Instead of stirring, the sample was replenished in the focus through a custom-built flow cell fitted with a syringe pump. 

\subsection {BOG transient-absorption controls}

\begin{figure*}[ht]
\centering
\includegraphics[width=0.7\textwidth,height=170mm,keepaspectratio]{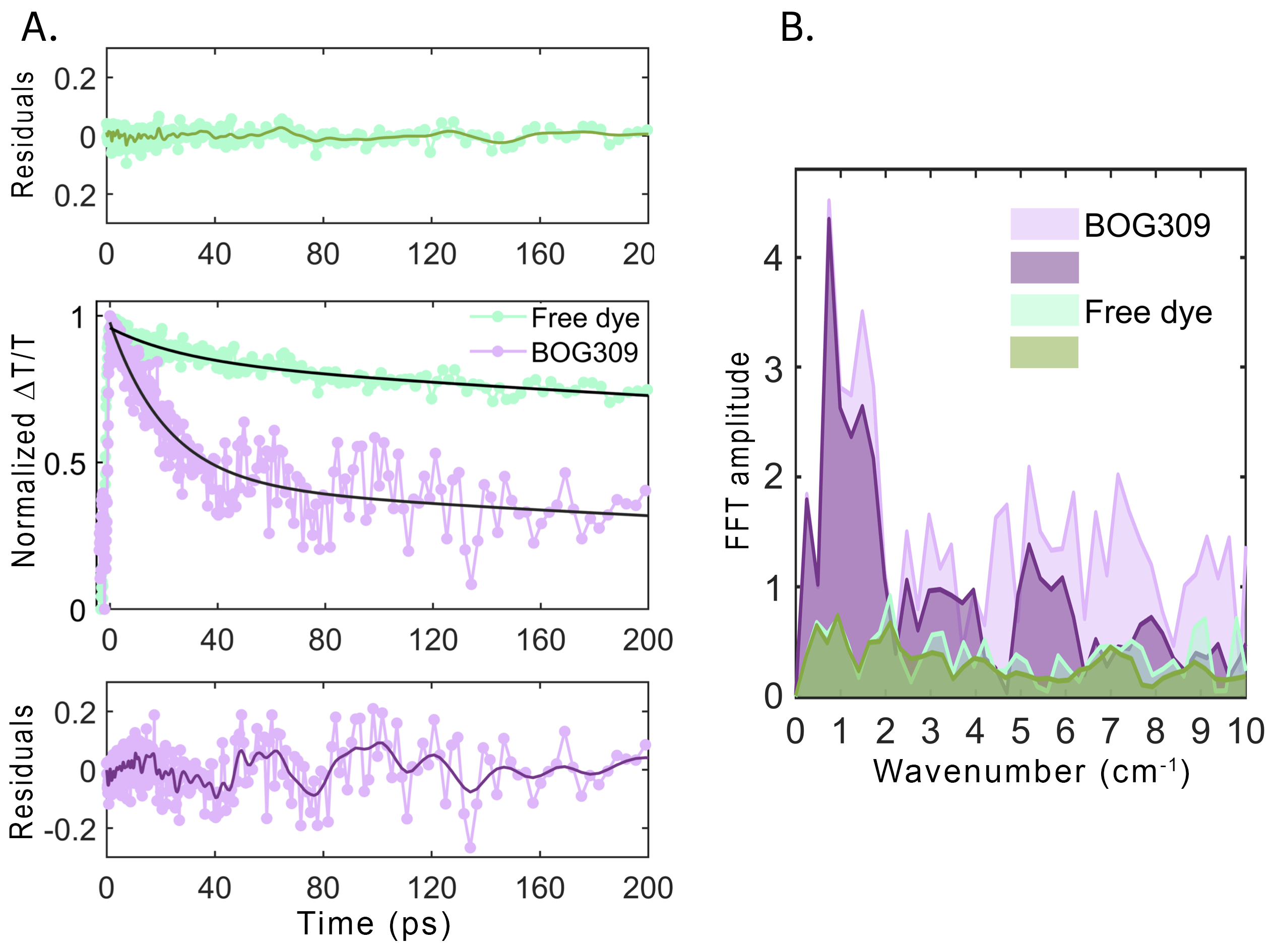}
\caption{A. Transient transmission traces of solutions of free OG (green) and BOG309 (purple, $<N>=309$) and the corresponding residuals after subtraction of biexponential fit. B. Amplitudes of the Fast Fourier transform of residuals for BOG309 and free dye from different samples, acquired in different days.}
\label{S:BOG309TA}
\end{figure*}

TA measurements on BOG solutions show distinctly different relaxation dynamics from free dyes solutions with the ground state bleach recovering almost completely within 35 ps from the excitation pulse due to SR damping, Appendix Fig.~\ref{S:BOG309TA}A. Results from different days and different samples show reproducible spectral features in the 1-2 wavenumber range, Appendix Fig.~\ref{S:BOG309TA}B..

\subsection{AuVLP transient-absorption controls}
\label{SI:TAAuVLP}

\begin{figure*}[!ht]
\centering
\includegraphics[width=0.55\textwidth,height=170mm,keepaspectratio]{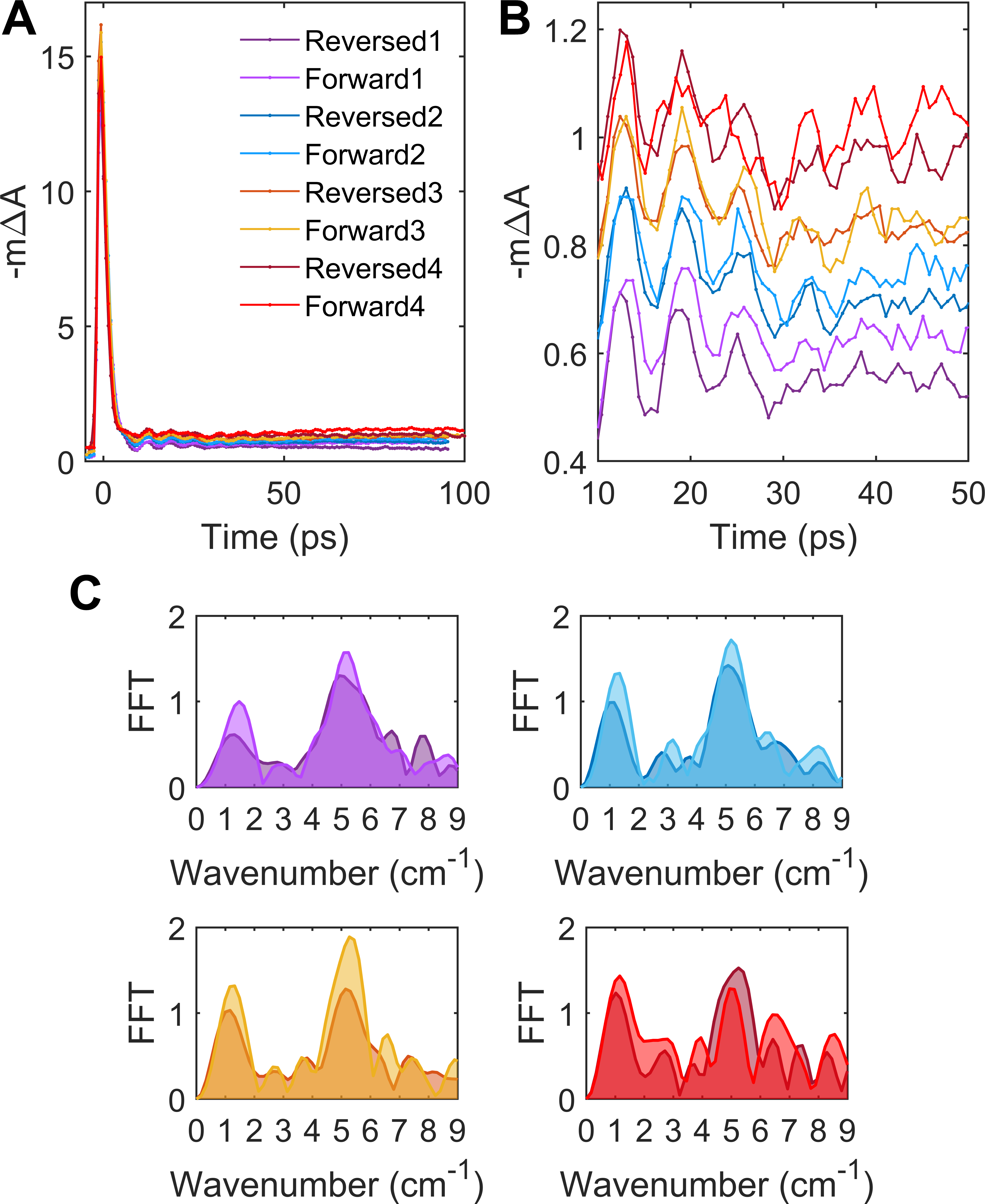}
\caption{\textbf{Transient absorption (TA) traces of AuVLP collected in one experiment}
		(A) Four pairs of TA traces, with label 1 denoting the first pair. Each pair consists of one forward scan and one reverse scan. (B) Enlarged view of Panel A. Colors correspond to the same legend as in Panel A. (C) Fourier transforms of four pairs of detrended and cropped traces (10–50 ps) from Panel B. Colors correspond to the legend in Panel A.}
\label{auvlprep}
\end{figure*}

Because slow baseline changes can arise from sample aggregation or delay-stage drift during extended scans, we tested whether the low-frequency modulation could be an acquisition artifact. Delay scans were acquired in alternating directions, with half recorded in the forward direction and half in the reverse direction. The \(\approx 1.3~\mathrm{cm^{-1}}\) modulation persisted in both scan directions, which argues against monotonic sample degradation or baseline drift as its origin, Appendix Fig.~\ref{auvlprep}. The feature also persisted over the pump-power range used in these measurements, arguing against a nonlinear artifact or damage-induced response. We therefore assign the low-frequency component to a capsid-associated acoustic mode mechanically coupled to the AuNP breathing coordinate.

\subsection{AuVLP data processing}
\begin{figure}[!ht]
\centering
\includegraphics[width=0.5\textwidth,height=170mm,keepaspectratio]{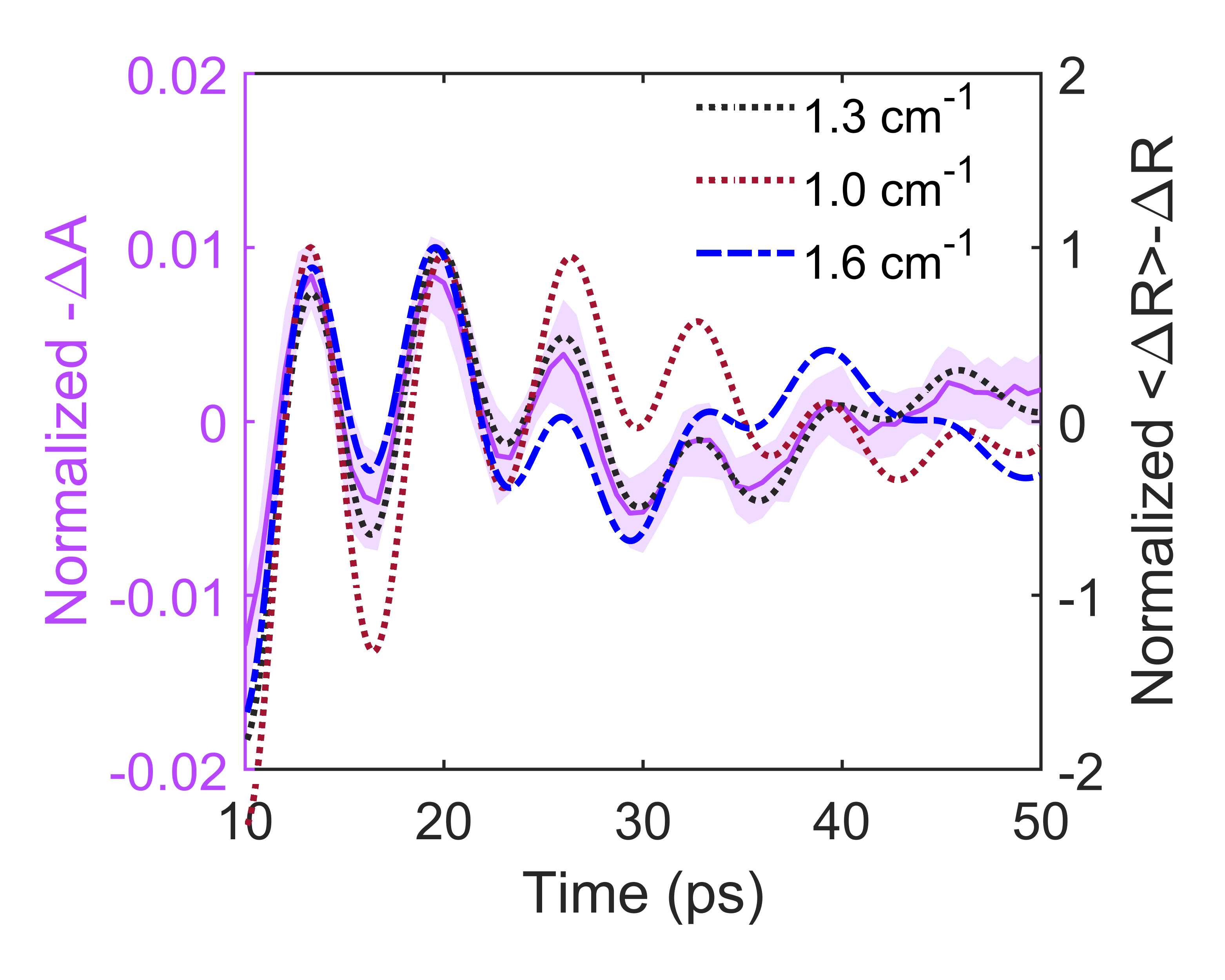}
\caption{\textbf{Comparing coupling constants in the model.} Purple: detrended average experimental TA trace (Parameters correspond to different $\sqrt{\kappa/m_2}$ values).}
\label{fit}
\end{figure}

Since $<\Delta R>-\Delta R$ is proportional to $-\Delta A$, the two are plotted together in Appendix Fig.~\ref{modelres}, for optimal fit parameters. In Appendix Fig.~\ref{modelres}A, the damping constant of the AuNP, $\gamma_{1}$, was first determined by fitting its experimental signal. For the AuVLP, the coupling constant $\kappa$ was initially estimated by assigning the resonance $\sqrt{\kappa/m_{2}}$ to the low-frequency FFT peak at $1.3~\mathrm{cm}^{-1}$. With this choice, the damping constant of the capsid, $\gamma_{2}$, was then obtained by optimally fitting the $\Delta R$ dynamics to the experimental TA trace (Appendix Fig.~\ref{modelres}B). Model parameters from fitting are provided in Appendix Table~\ref{edt:core}b. Note that the frequency of $1.3~\mathrm{cm}^{-1}$ gives the best agreement with the experimental signal in both the time domain and the Fourier transform (Appendix Fig.~\ref{modelres}C), while values of $1.0$ or $1.6~\mathrm{cm}^{-1}$ show poorer agreement  (Appendix Fig.~\ref{fit}). The slow modulation at $1.3~\mathrm{cm}^{-1}$ arises from one of the two (weakly) coupled eigenmodes,  rather than a beating between two modes\cite{Frimmer2014}. Also note that, by using the coupled one-dimensional oscillators model we are not suggesting that only the breathing mode of the capsid should be taken into consideration. Several low-frequency modes of the capsid are likely to be excited at the same time, each with their own damping (see below, Appendix Fig.~\ref{capsid_normalmodes} ). However, our experiment cannot resolve them. 

\begin{figure*}[!ht]
\centering
\includegraphics[width=0.6\textwidth,height=170mm,keepaspectratio]{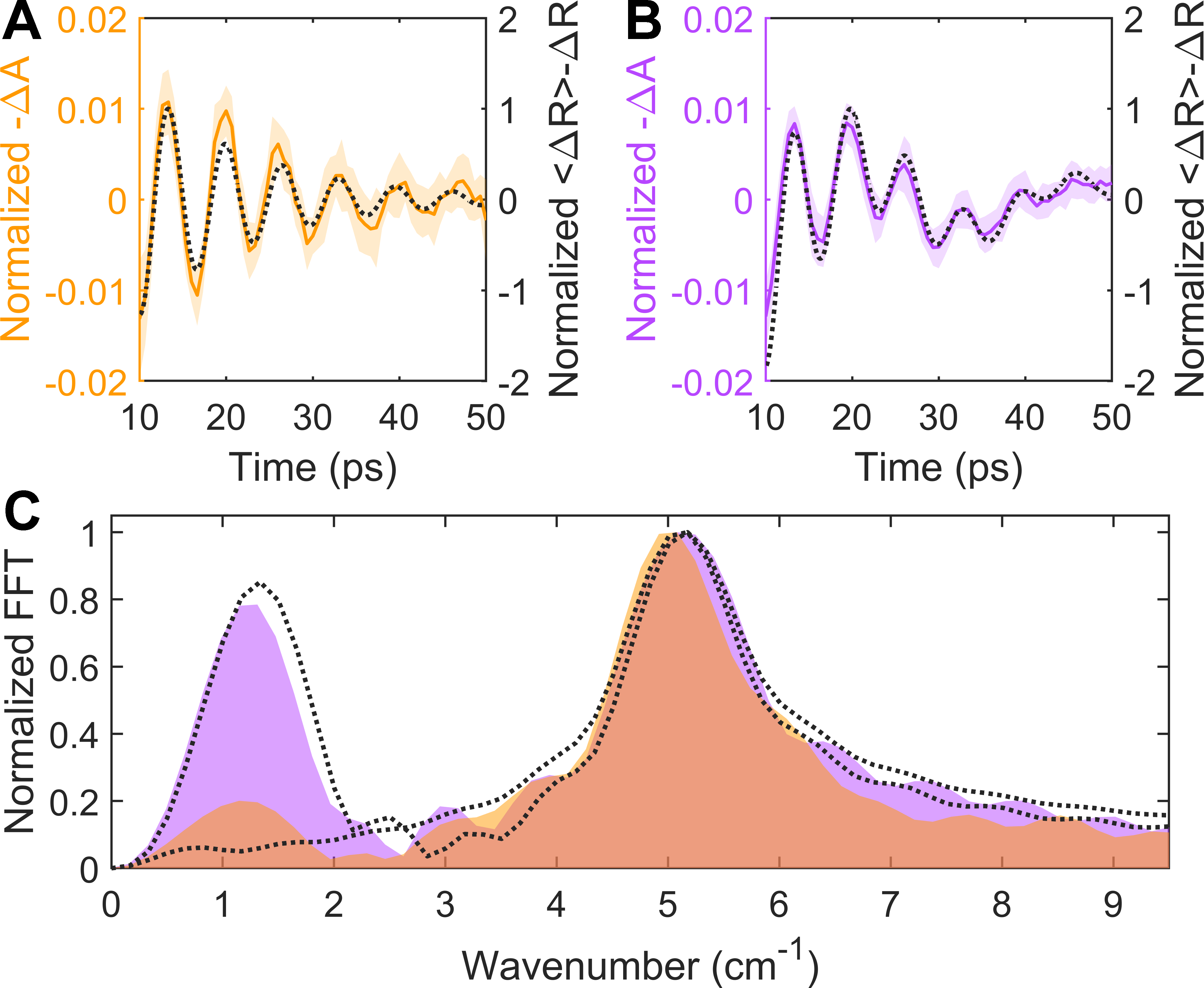}
\caption{\textbf{Experimental transient absorption (TA) trace versus modelled radius change} (A) AuNP. The colored line shows the oscillation in the detrended average TA trace, and the black dashed line represents the radius change variation from the model. 
(B) AuVLP. Same representation as in panel~A. 
(C) Fourier transforms of the TA (yellow: AuNP; purple: AuVLP) and the corresponding modelled radius change variations (black dashed). Spectra were normalized to unity at the maximum. Model parameters are listed in Appendix Table~\ref{edt:core}b.
 }
\label{modelres}
\end{figure*}

\subsection{Harmonic inversion analysis}

For an alternative approach without detrending, we fed 40 raw experimental traces into Harminv v1.4.2 (filter--diagonalization method)\cite{MandelshtamTaylor1997,harminv-1.4.2}. From the resolved modes, we retained only those with \(Q>0\) and a reported relative error \(\varepsilon\) smaller than \(1/\)decay rate for plotting in Appendix Fig.~\ref{harminv}. The results again show a peak around \(5~\mathrm{cm^{-1}}\) for AuNP breathing, and a second low-frequency peak (slightly above \(1~\mathrm{cm^{-1}}\)) for the capsid phonon of AuVLP. 

\begin{figure}[!ht]
\centering
\includegraphics[width=0.5\textwidth,height=170mm,keepaspectratio]{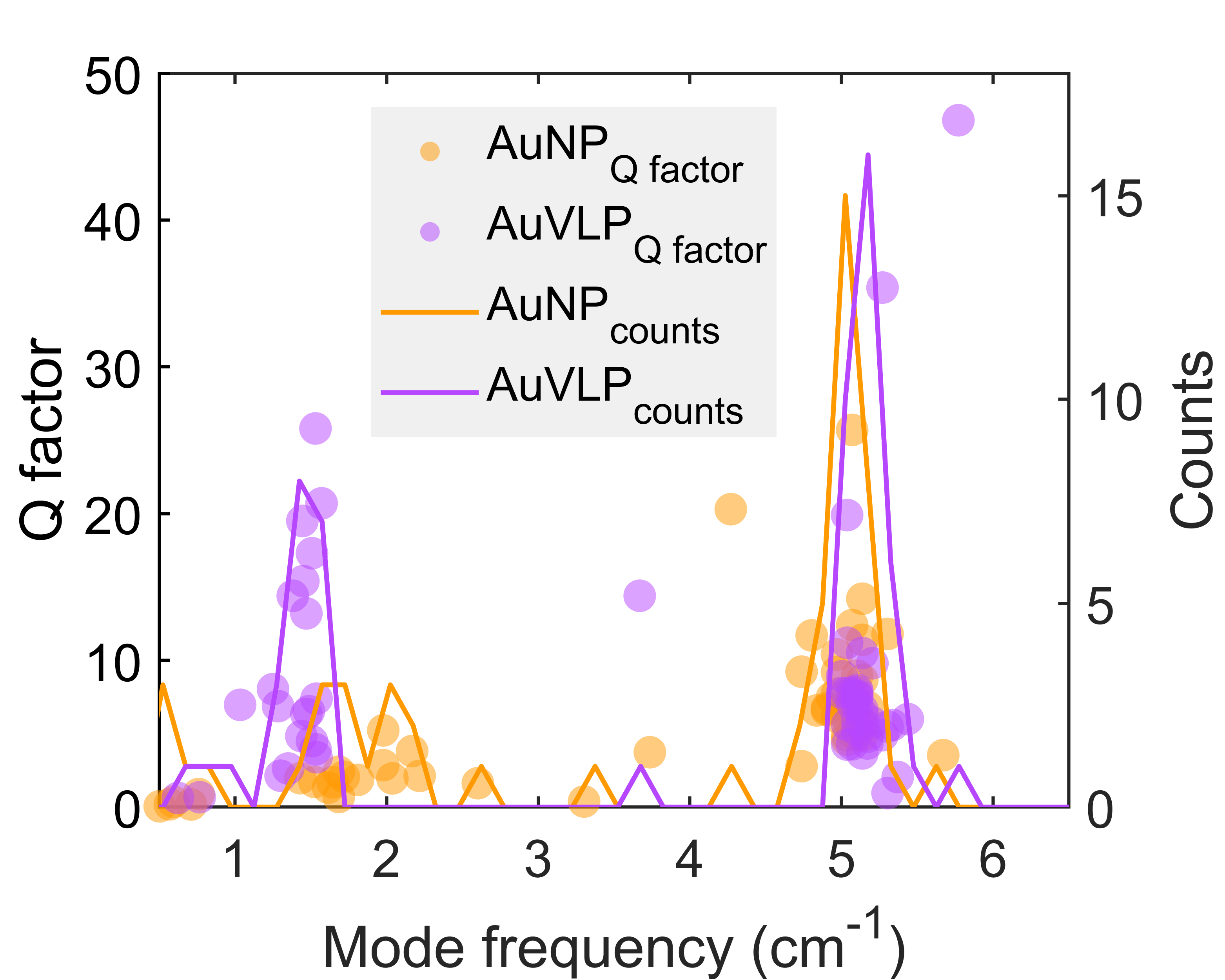}
\caption{\textbf{Harmonic Inversion of Time Signals analysis of AuNP/AuVLP.}
		The scatter points (left y-axis) show mode Q factors; the line plots (right y-axis) display the histogram counts of mode populations\cite{MandelshtamTaylor1997,harminv-1.4.2}.}
\label{harminv}
\end{figure}

\newpage

\subsection{Capsid vibrational calculations}\label{methods:capsid}We developed a coarse-grained elastic network model\cite{tirion1996large}, which we combined with a Langevin friction model\cite{lamm1986langevin,miller2008langevin}, to describe the equilibrium vibrational dynamics of the capsid in solution. Our coarse-grained representation of the capsid involves a single mass particle with mass 20295Da for each of the 180 coat proteins of the brome mosaic virus. We constructed the equilibrium configuration of the capsid by placing the centers of 12 pentamers at the vertices of a regular icosahedron with 12.092nm edge length (the radius of the circumscribed sphere is 11.5nm, which equals the mid-radius of the brome mosaic virion) and 20 hexamers at the centers of the icosahedron faces (the centers of the faces lie at the vertices of the circumscribed dodecahedron). We expanded each pentamer with 5 coat protein mass particles, which we first positioned at 2.95nm away from the pentamer center in direction to one of the neighboring hexamers and then rotated the five particles at $30^{\circ}$ around the virion radius vector piercing through the pentamer center. We similarly expanded each hexamer with 6 coat protein mass particles, which we first positioned at 2.95nm away from the hexamer center in direction to one of the three neighboring pentamers and three neighboring hexamers and then rotated the six particles at $30^{\circ}$ around the virion radius vector piercing through the hexamer center. The resulting capsid equilibrium configuration shown in Appendix Fig.~\ref{capsid_normalmodes}A incorporates the T=3 icosahedral symmetry of the brome mosaic virus and resembles the virion crystal structure. 

\begin{figure*}[!ht]
\centering
\includegraphics[width=0.85\textwidth,height=170mm,keepaspectratio]{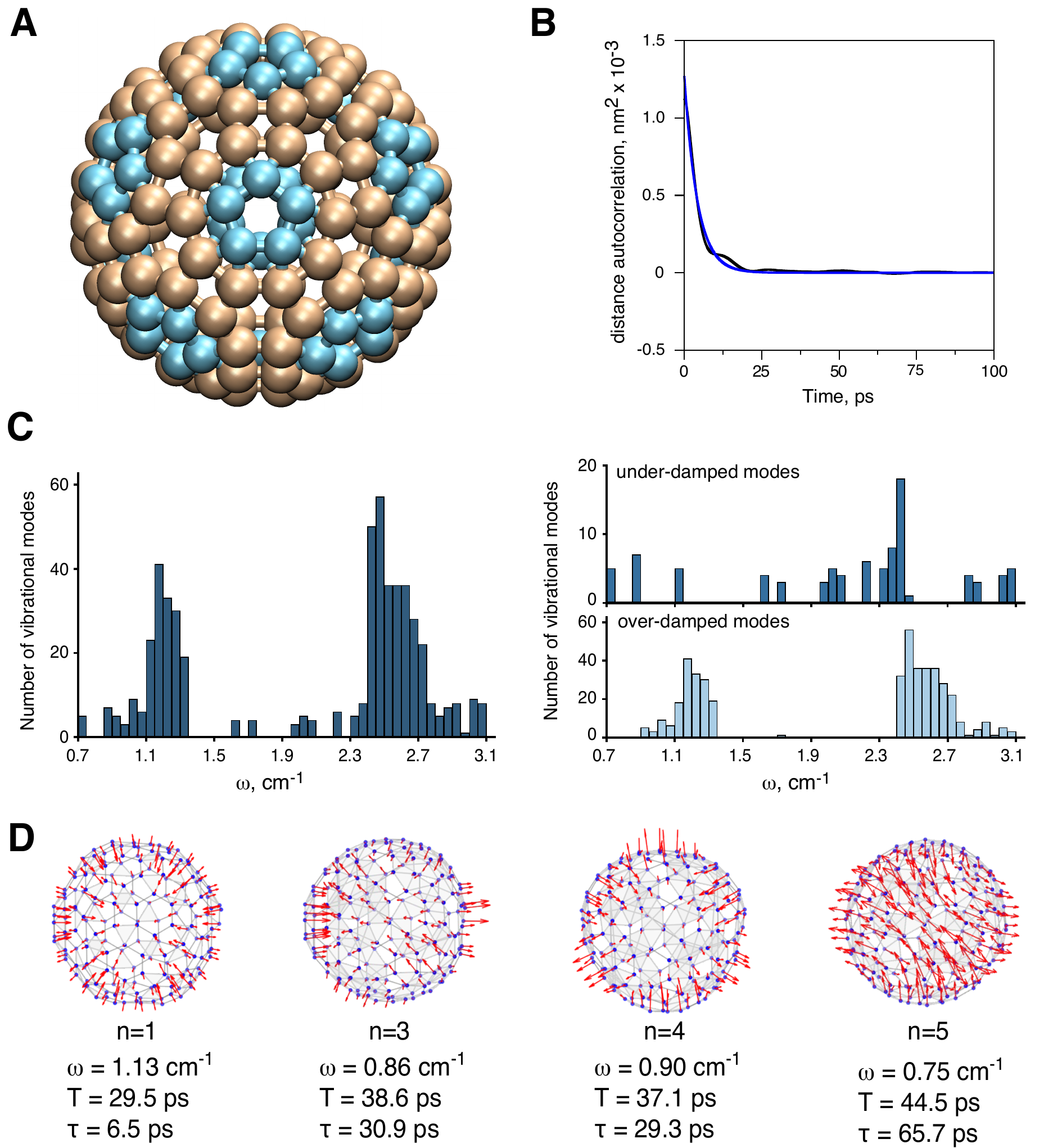}
\caption{Capsid vibrational dynamics.
		(A) Coarse-grained model of the capsid. Spheres depict the coat protein particles, pentamers are in blue, and hexamers are in orange. (B) Distance between nearest neighbor coat proteins autocorrelation function (black line) together with an exponential fit (blue line). (C)  Density of vibrational modes from the elastic-network model (left) together with density of under-damped (right above) and over-damped (right below) modes. Under-damped modes have a relaxation time that is commensurate or longer than their period, whereas over-damped modes have a relaxation time that is much shorter than their period. (D) The lowest energy mode from each degenerate group is shown along with its frequency $\omega$, period T, and relaxation time $\tau$. The red arrows depict the normal mode displacements with blue dots the viral coat proteins.}
\label{capsid_normalmodes}
\end{figure*}

To compute the capsid Hessian matrix and the associated vibrational modes and frequencies, we used an elastic network model with the potential energy\cite{tirion1996large}
\begin{equation}
U = \frac{K}{2} \sum_{i \neq j} (r_{ij} - r^0_{ij})^2,
\end{equation}
where $i$ and $j$ run over coat protein particles that are within $r_c = 10\text{nm}$ apart, $r_{ij}$ is the distance between the proteins, $r^0_{ij}$ is the equilibrium distance between them based on the equilibrium capsid configuration, and $K=24073.5 \text{cm}^{-1}/\text{nm}^2$ is the elastic spring constant. Both parameters of the model, $K$ and $r_c$, are adjusted to reproduce the low frequency vibrations of the brome mosaic virion from atomistic simulations\cite{dykeman2010atomistic}.

We developed the capsid friction model using the Rotne-Prager hydrodynamic interaction tensor\cite{lamm1986langevin}, $T_{ij}$, which takes into account the shielding of friction forces due to the overlap of the coat protein hydrodynamic spheres
\begin{equation}
\begin{split}
T_{ij} &= 1, \; \text{when} \; i=j\\
T_{ij} &= \frac{3a}{4r_{ij}} \left[ 1 + \frac{\vec{r}_{ij} \otimes \vec{r}_{ij}}{r^2_{ij}} \right] + \frac{3a^3}{2r^3_{ij}} \left[ \frac{1}{3} - \frac{\vec{r}_{ij} \otimes \vec{r}_{ij}}{r^2_{ij}} \right], \; \text{when} \; i \neq j, \; r_{ij} \ge 2 a\\
T_{ij} &= \left[ \left(1 - \frac{9 r_{ij}}{32a} \right) + \frac{3}{32} \frac{\vec{r}_{ij} \otimes \vec{r}_{ij}}{ar_{ij}} \right], \; \text{when} \; i \neq j, \; r_{ij} < 2 a,
\end{split}
\end{equation}
where $r_{ij}$ is the distance between the coat proteins, and $a=1.815 \text{nm}$ is the coat protein hydrodynamic radius. We computed the mass-weighted friction matrix, $\gamma_{ij}$, from the inverse of the hydrodynamic tensor, $\gamma_{ij} = \gamma T^{-1}_{ij}$, where $\gamma=0.193 \text{ps}^{-1}$ is the relaxation rate of the solvent. Both $\gamma$ and $a$ are adjusted to reproduce previously computed relaxation times of the capsid eigenmodes\cite{dykeman2010atomistic}.

To calculate the undamped normal modes and vibrational frequencies of the capsid, we diagonalized the mass-weighted elastic network Hessian, $\boldsymbol{H}/m$. Similarly, to compute the damped vibrational modes and mode relaxation times, we diagonalized the non-symmetric matrix $\boldsymbol{A}$
\begin{equation}
\boldsymbol{A} = 
\begin{pmatrix}
  \boldsymbol{0} & \boldsymbol{1}\\
  -\boldsymbol{H}/m & -\boldsymbol{\gamma}
\end{pmatrix},
\end{equation}
where $\boldsymbol{H}$ is the elastic network Hessian matrix, and $\gamma$ is the mass-weighted friction matrix evaluated at the equilibrium capsid configuration. The eigenvalues of the $\boldsymbol{A}$-matrix are generally complex with negative real parts whose absolute values equal the inverse mode relaxation times. We assigned mode relaxation times to the undamped vibrational modes of the capsid by establishing pairs of corresponding modes between the undamped and damped normal modes based on mode overlap scores. We plot in Appendix Fig.~\ref{capsid_normalmodes} the mode periods and relaxation times together with the mode degeneracies.

We simulated the protein-protein distance autocorrelation function, $\langle r_{ij}(t) r_{ij}(0) \rangle$, for nearest neighbor proteins $i$ and $j$ to characterize the vibrational dynamics of the local molecular environment of interstitially attached chromophores at equilibrium. To compute the correlation function, we generated a 1.31ns molecular dynamics trajectory by solving the Langevin equation of motion
\begin{equation}
    m \ddot{r}_{i\alpha} + m \sum_{j\beta} \gamma_{i\alpha, j\beta} \dot{r}_{j \beta} + \sum_{j\beta} H_{i\alpha, j\beta} (r_{j \beta}-r^0_{j \beta}) = R_{i\alpha}(t),
\end{equation}
where $r_{i\alpha}$ are the $i^{\text{th}}$ coat protein coordinates with $\alpha$ the Cartesian components, $r^0_{i\alpha}$ are the equilibrium coat protein positions, $\gamma_{i\alpha, j\beta}$ is the mass-weighted friction matrix, $H_{i\alpha, j\beta}$ is the elastic network Hessian, and $R_{i\alpha}$ are uncorrelated random numbers with zero mean and variance, satisfying the fluctuation-dissipation theorem, $\langle R_{i\alpha} R_{j\beta} \rangle=2m\gamma_{i\alpha, j\beta}/k_BT$. We used a quasi-simplectic stochastic leap-frog integrator\cite{mannella2004quasisymplectic,mannella2006numerical} with a timestep of 1fs to solve the equations of motion. We initially equilibrated the capsid for 50 ps, and sampled the nearest neighbor coat protein distances every 10 steps from the following 1.31 ns molecular dynamics trajectory to calculate the correlation function. The resulting distance correlation function at $T=300$ K is shown in Appendix Fig.~\ref{capsid_normalmodes}B together with an exponential fit in blue. The zero time value of the correlation function provides the root-mean-square deviation of the nearest-neighbor coat protein distances of 0.035 nm and the relaxation time $1/\eta$ of 4 ps characterizes the timescale of stochastic fluctuations. As expected at thermal equilibrium, the correlation function shows a nearly smooth exponential behavior that is characteristic of over-damped dynamics with a few oscillations, originating from under-damped, slowly relaxing modes. 

Vibrational spectra and the classification of modes are discussed in Supplementary Note~\ref{SI:spectrum}.

\subsection{Chromophore electronic structure}\label{methods:chromophore}
To elucidate how the molecular geometry of the chromophores changes upon electronic excitation and to gain insight into how the chromophores couple to the capsid vibrational motion, we computed the optimal ground- and first excited-state molecular geometries of the dye. The ground-state geometry is representative of the molecular configurations where absorption occurs, whereas the excited-state geometry is representative of the molecular configurations where fluorescence occurs. Upon electronic excitation, dye molecules with configurations around the optimal ground-state geometry are promoted to the first electronically excited state, and in the subsequent internal conversion they move away toward configurations around the optimal excited-state geometry. We optimized the geometry of the dianion form of Oregon Green 488 using Density Functional Theory (DFT) as implemented in ORCA (v. 6.0.1) with the CAM-B3LYP functional, the ma-def2-SVP basis set, and the RIJCOSX resolution of identity \cite{RN178,RN28,RN205,RN21,RN29,RN56}. We accounted for dispersion interactions using the D3 dispersion correction with Becke-Johnson damping (D3BJ)\cite{RN214,RN215}, and we incorporated the dielectric environment of the viral capsid using the conductor-like polarizable continuum model (CPCM) with an epsilon value of 37.5, which is characteristic for the solvent acetonitrile\cite{RN167}. We employed time-dependent density functional theory (TD-DFT) to calculate the first electronic excitation energy and to optimize the first excited-state geometry using the same functional, basis set, and dispersion and solvent corrections as in the ground-state calculations. We computed how the transition energy depends on the molecular rearrangement by interpolating the geometry between the ground- and excited-state configurations at steps of 10\%. The results in Appendix Fig.~\ref{og488_tddft} show that upon electronic excitation the dihedral angle between the xanthene core and the benzoic acid moiety of Oregon Green 488 increases by $16^\circ$ (Appendix Fig.~\ref{og488_tddft}A). This change of the dihedral angle is driven by the delocalization of the lowest unoccupied molecular orbital (LUMO), which accepts an electron upon excitation, to the benzoic acid moiety (Appendix Fig.~\ref{og488_tddft}B). Furthermore, the calculations of the transition energy along the path between the ground- and excited-state configurations show that the optical transition energy shifts by $0.145\text{eV}$ upon the dihedral angle distortion (Appendix Fig.~\ref{og488_tddft}C), revealing a strong electron-vibrational coupling between the electronic transition and the relative orientation of the xanthene core and the benzoic acid moiety. These results suggest that the optical transition of dye molecules, which are bound in the confined space between capsid coat proteins, may couple to the capsid vibrational motion via modulation of the dye dihedral angle caused by fluctuations of the inter-capsid separation. We incorporate this electron-vibrational coupling in our minimal model of vibrationally-assisted superradiance in Appendix~\ref{VibAssSR}.

\begin{figure*}[!ht]
\centering
\includegraphics[width=1\textwidth,height=170mm,keepaspectratio]{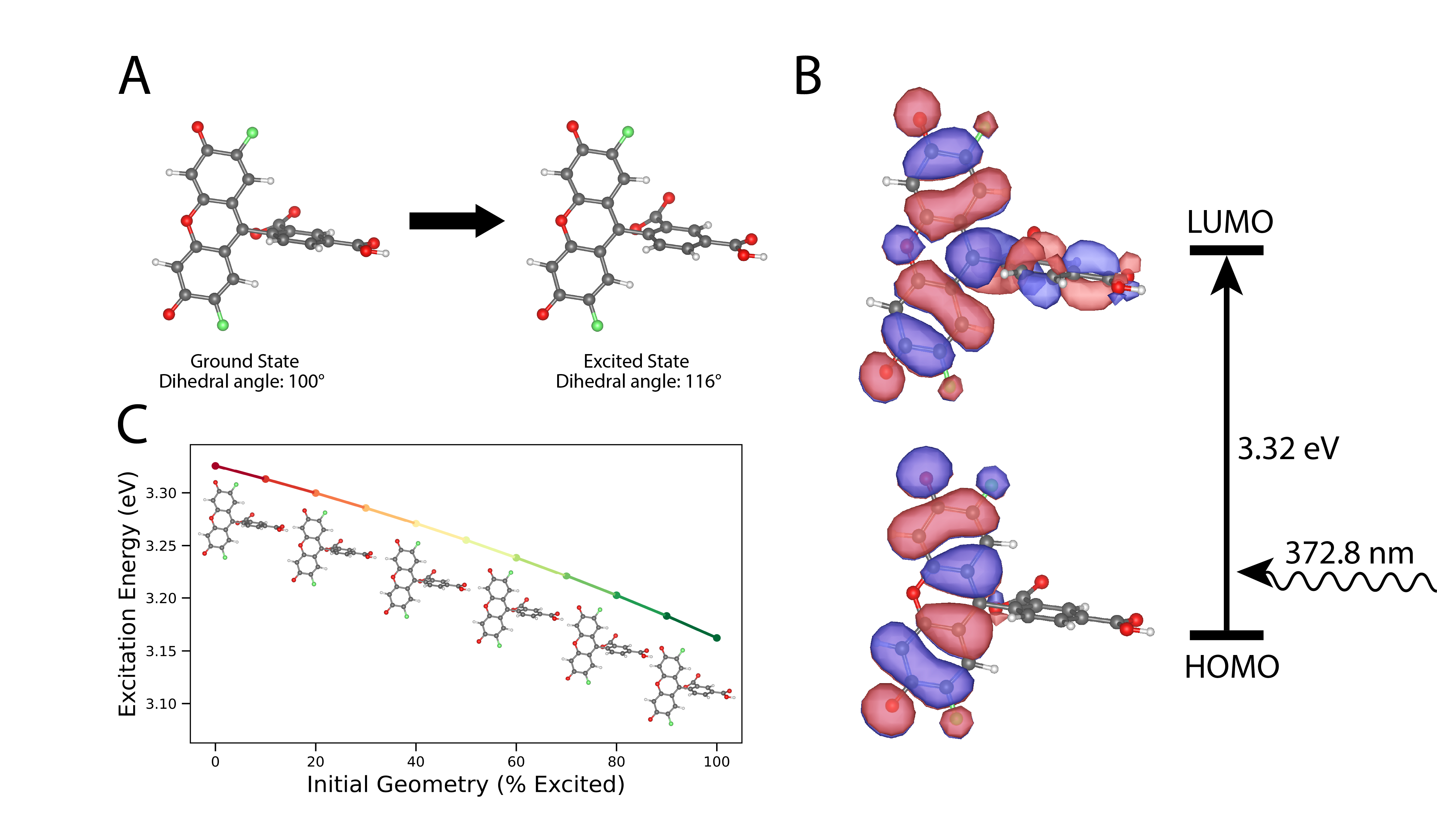}
\caption{\textbf{TD-DFT excited state transition analysis of Oregon Green 488.}
		(A) The dihedral angle between the xanthene core and benzoic acid portion of Oregon Green 488 differs by 16° by DFT geometry optimization. (B) The first excited state transition consists almost entirely of a HOMO to LUMO transition. The HOMO orbital shows electron density localized on the xanthene core, while the LUMO orbital shows a transfer of electron density to the benzoic acid substituent. The transition energy between the two states is calculated to be 3.32 eV (372.8 nm) at the ground state optimized geometry. (C) The first excited-state transition energy decreases to 3.175 eV upon reorganization of the molecular geometry to the optimal excited state geometry, implying a Stokes shift of 0.145 eV. The variation of the transition energy between the optimal ground-state and optimal excited-state geometries is nearly linear.}
\label{og488_tddft}
\end{figure*}

\subsection{Collective-emission model}
The open-system model and the reduced two-chromophore analysis are described in Supplementary Note~\ref{SI:Qmodel}. The model parameters are listed in Appendix Table~\ref{edt:models}a.

The accompanying background schematics are in Appendix Figs.~\ref{cartSR} and~\ref{overv}. The core--shell mechanical model is illustrated in Appendix Fig.~\ref{spring}; its analytical development is given in Supplementary Note~\ref{SI:CoupledOscillators}. Continuum-mode estimates and the constants used are in Appendix Table~\ref{edt:models}b,c, with derivations in Supplementary Note~\ref{SI:Identification}.

\begin{figure}[!ht]
\centering
\includegraphics[width=0.25\textwidth,height=170mm,keepaspectratio]{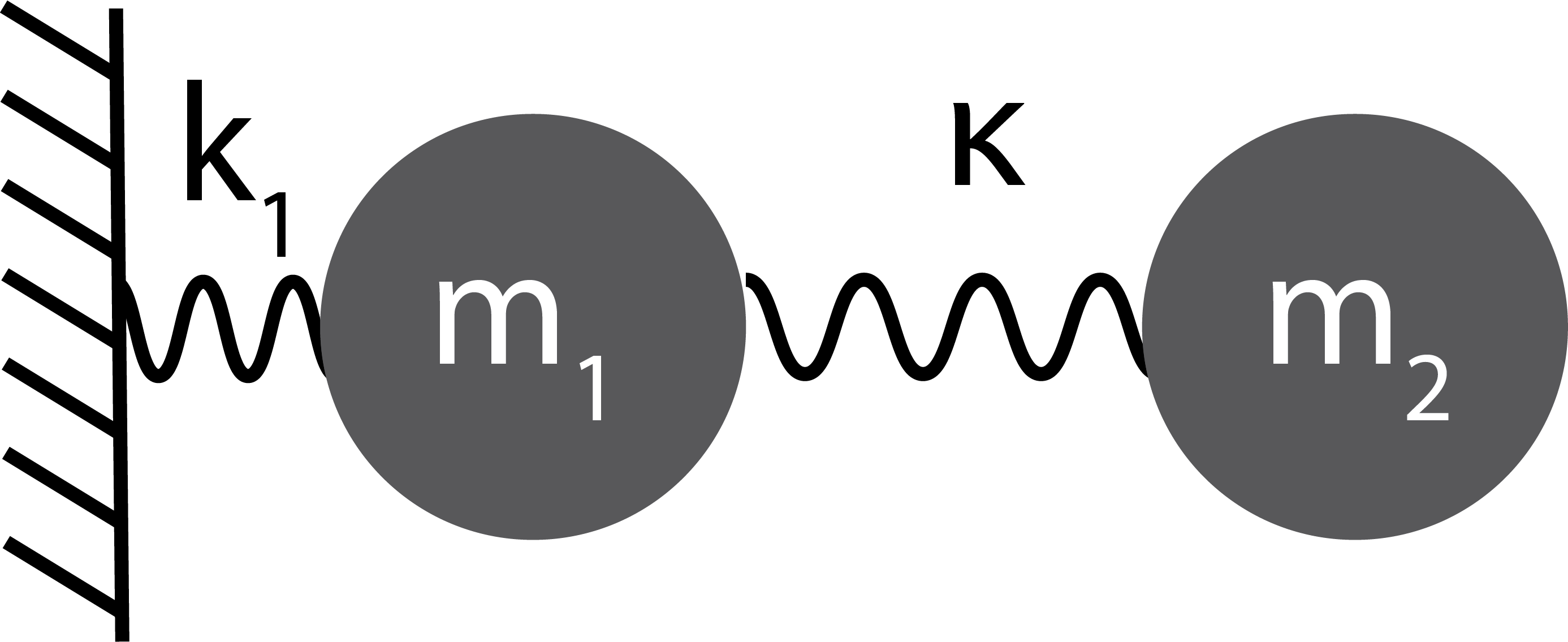}
\caption{Toy core-shell vibrational model: Two damped oscillators with masses m$_1$ and m$_2$ and two springs with force constants k$_1$ (for the Au NP lattice) and $\kappa$ (for the coat-protein shell and NP coupling).}
\label{spring}
\end{figure}

\clearpage
\begin{table*}[!ht]
\caption{\textbf{Parameters and estimates for theoretical models.}}
\label{edt:models}
\begingroup\setstretch{1.15}\small
\noindent \textbf{a} \textbf{Superradiance model Hamiltonian parameters.}\par\smallskip
\begin{center}\begin{adjustbox}{max width=\textwidth}\begin{tabular}{l l l}
        \hline
        \textbf{Parameter} & \textbf{Description} & \textbf{Value} \\
        \hline
        $1/\eta$ & OU relaxation time & 4 ps \\
        $\Delta$ & OU broadening & $2 \pi \times 10$ THz \\
        $\Omega_{EE}$ & Exciton interaction strength& $2 \pi \times 1.75$ THz \\
        $\Omega_{EPh}$ & Exciton-phonon coupling& $2 \pi \times 10$ THz \\
        $\Omega_{EEPh}$ & Exciton-interaction-phonon coupling& $2 \pi \times -0.1$ THz \\
        $\Omega_{Ph}$ & Phonon frequency & $2 \pi \times 0.035$ THz \\
        $1/\eta_{Ph}$ & Underdamped phonon relaxation time& 65.0 ps \\
        $1/\Gamma$ & Single chromophore lifetime & 200 ps \\
        \hline
    \end{tabular}\end{adjustbox}\end{center}

\noindent \textbf{b} Estimated acoustic phonon mode frequencies, $\omega$, and lifetimes, $\tau$, for the lowest spheroidal modes for AuVLP and BOG309 in water~\cite{Saviot2007Damping,saviot2007virus,saviot_cfm_calculator}\par\smallskip
\begin{center}\begin{adjustbox}{max width=\textwidth}\begin{tabular}{lccc}
\\
\hline
Sample &
\begin{tabular}[c]{@{}c@{}}$l=0$\\ $\omega,\tau$\\ (cm$^{-1}$, ps)\end{tabular} &
\begin{tabular}[c]{@{}c@{}}$l=1$\\ $\omega,\tau$\\ (cm$^{-1}$, ps)\end{tabular} &
\begin{tabular}[c]{@{}c@{}}$l=2$\\ $\omega,\tau$\\ (cm$^{-1}$, ps)\end{tabular} \\
\hline
AuVLP  & 1.27, 7.4 & 0.48, 43  & 0.52, 35 \\
BOG & 1.42, 7.4 & 0.42, 6.0 & 0.58, 36 \\
\hline
\end{tabular}\end{adjustbox}\end{center}

\noindent \textbf{c} \textbf{Material constants used for mode calculation in Appendix Table~\ref{edt:models}b.}~\cite{saviot2007virus}\par\smallskip
\begin{center}\begin{adjustbox}{max width=\textwidth}\begin{tabular}{lccccc}
\\
\hline
Material 
& \begin{tabular}[c]{@{}c@{}}Mass density\\ $\rho$ (g/cm$^{3}$)\end{tabular}
& \begin{tabular}[c]{@{}c@{}}Longitudinal sound\\ velocity $v_{L}$ (m/s)\end{tabular}
& \begin{tabular}[c]{@{}c@{}}Transverse sound\\ velocity $v_{T}$ (m/s)\end{tabular}
& \begin{tabular}[c]{@{}c@{}}Shear\\ viscosity $\mu$ (Pa$\cdot$s)\end{tabular}
& \begin{tabular}[c]{@{}c@{}}Bulk\\ viscosity $\zeta$ (Pa$\cdot$s)\end{tabular}
\\
\hline
Gold  & 19.32 & 3330 & 1250 & 0 & 0 \\
Water & 1.00  & 1498 & 0    & 0.00089 & 0.00215 \\
Virus & 1.20  & 1817 & 915  & 0 & 0 \\
\hline
\end{tabular}\end{adjustbox}\end{center}
\endgroup
\end{table*}

\clearpage
\section{Background on superradiance and brome mosaic virus}
\subsection{Superradiance}\label{SI:background}

\begin{figure*}[ht]
\centering
\includegraphics[width = \textwidth,height=170mm,keepaspectratio]{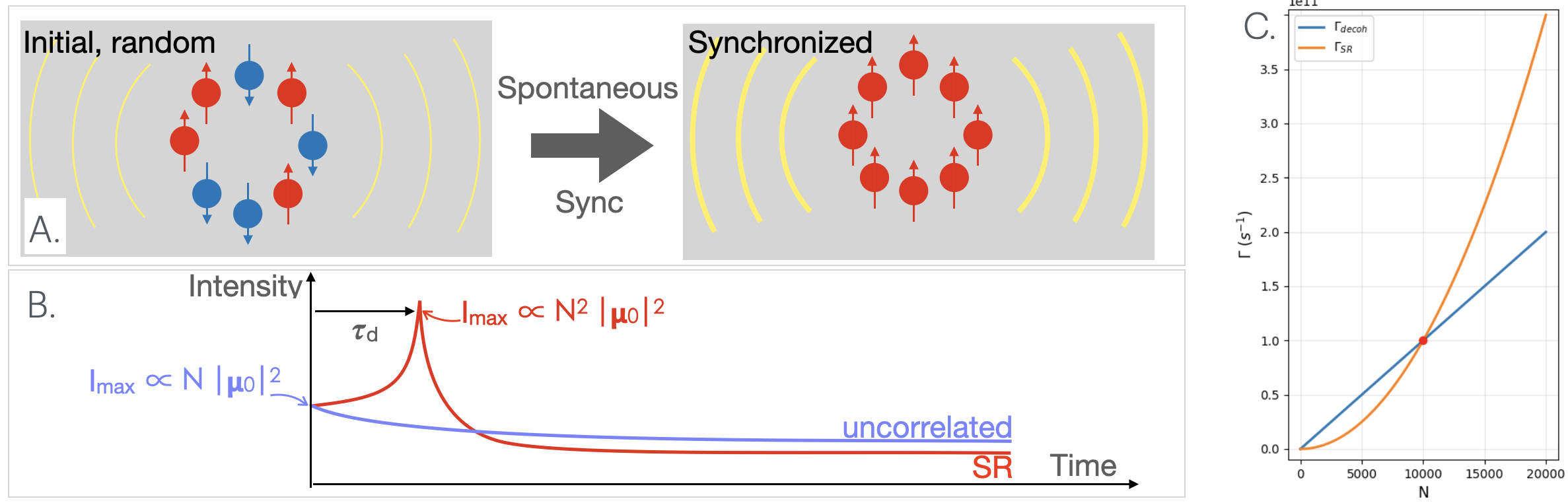}
\caption{A. Cartoon of spontaneous transition from a state of random phases to a state of identical  phases of an ensemble of emitting dipoles. B. Sketch of resulting change from uncorrelated to correlated emission. $\tau_d$ is the delay time between excitation and the formation of a fully coherent state. C. For large emitter ensembles, the collective emission rate ($\propto N^2$) wins over the collisional decoherence rate ($\propto N$).}
\label{cartSR}
\end{figure*}

Superradiance (SR) is a quantum optical phenomenon that arises when $N$ indistinguishable quantum emitters interact with a common electromagnetic field and collectively radiate as a single coherent entity \cite{Dicke1954, Benedict1996}. For subwavelength assemblies ($L \ll \lambda$) in the fully symmetric collective state, the dipoles add coherently to produce an effective dipole moment:

\[
\vec{P}=N\vec{\mu}_0,
\]

where $\vec{\mu}_0$ is the transition dipole moment of an individual emitter.

Because the spontaneous emission rate scales with the square of the transition dipole moment,

\[
\gamma = \frac{4}{3}\frac{|\vec{P}|^2\omega_0^3}{\hbar c^3},
\]

cooperative emission produces an enhanced radiative decay rate

\[
\Gamma_{SR}\sim \frac{N}{\tau_{sp}},
\]

and a correspondingly shortened emission duration

\[
\tau_{SR}\propto \frac{\tau_{sp}}{N}.
\]

The peak emission intensity therefore scales approximately as

\[
I_{SR}^{max}\propto N^2.
\]

Formation of the collective state requires a finite synchronization time. For an ideal ensemble of two-level emitters at 0 K, the delay between excitation and the onset of the SR burst is approximately

\[
\tau_d \sim \tau_{sp}\frac{\log N}{N},
\]

which decreases slowly with increasing emitter number \cite{Gross1982a}.

\subsection{The brome mosaic virus}

Brome mosaic virus (BMV) is a small, non-enveloped, positive-sense single-stranded RNA plant virus with a T=3 icosahedral capsid with an outer diameter of approximately 28 nm, an internal cavity of approximately 18 nm, and a protein shell about 5 nm thick\cite{Lucas2002}. The capsid interior is positively charged, due to flexible N-terminal arm of each coat protein containing seven arginines and one lysine that extend toward and electrostatically associate with the negatively charged RNA genome. These capsid--RNA contacts contribute to genome encapsidation, particle integrity, and mechanical stiffness; genome packing has also been associated with an effective negative internal pressure of approximately 0.1 atm that prestresses and stiffens the shell.

Each BMV coat protein contains 12 lysines - 2160 nominal amines per virion - but steric accessibility and local interactions restrict Oregon Green 488 NHS-ester labeling to only a subset, with intact particles typically saturating near 300 dyes per virion. The lysines are structurally assigned as internal residues, external residues, and intersubunit. Proteomic measurements identify the principal OG-reactive population as K8, K41, K44, K86, K64 and K105, with lower occupancy at K81 and 165 and no OG labeling at K53, K83, K111, or K130\cite{AnilSushma2021}. The external K64/K105/K165 sites are separated by approximately 11--35\r{A} and K86 lies only approximately 11.3--11.9\r{A}  from K81/K83 and an OG dimension of roughly 10\r{A}.  These distances permit strong F{\"o}rster-type interactions\cite{Li2026a} but prevent simultaneous occupation of every lysine, producing an intrinsically heterogeneous dye distribution. From dye saturation amount, estimated average nearest-neighbor separation among candidate labeling sites is about 1.9 nm.

\begin{figure}[t]
\centering
\includegraphics[width = 0.5 \textwidth,height=170mm,keepaspectratio]{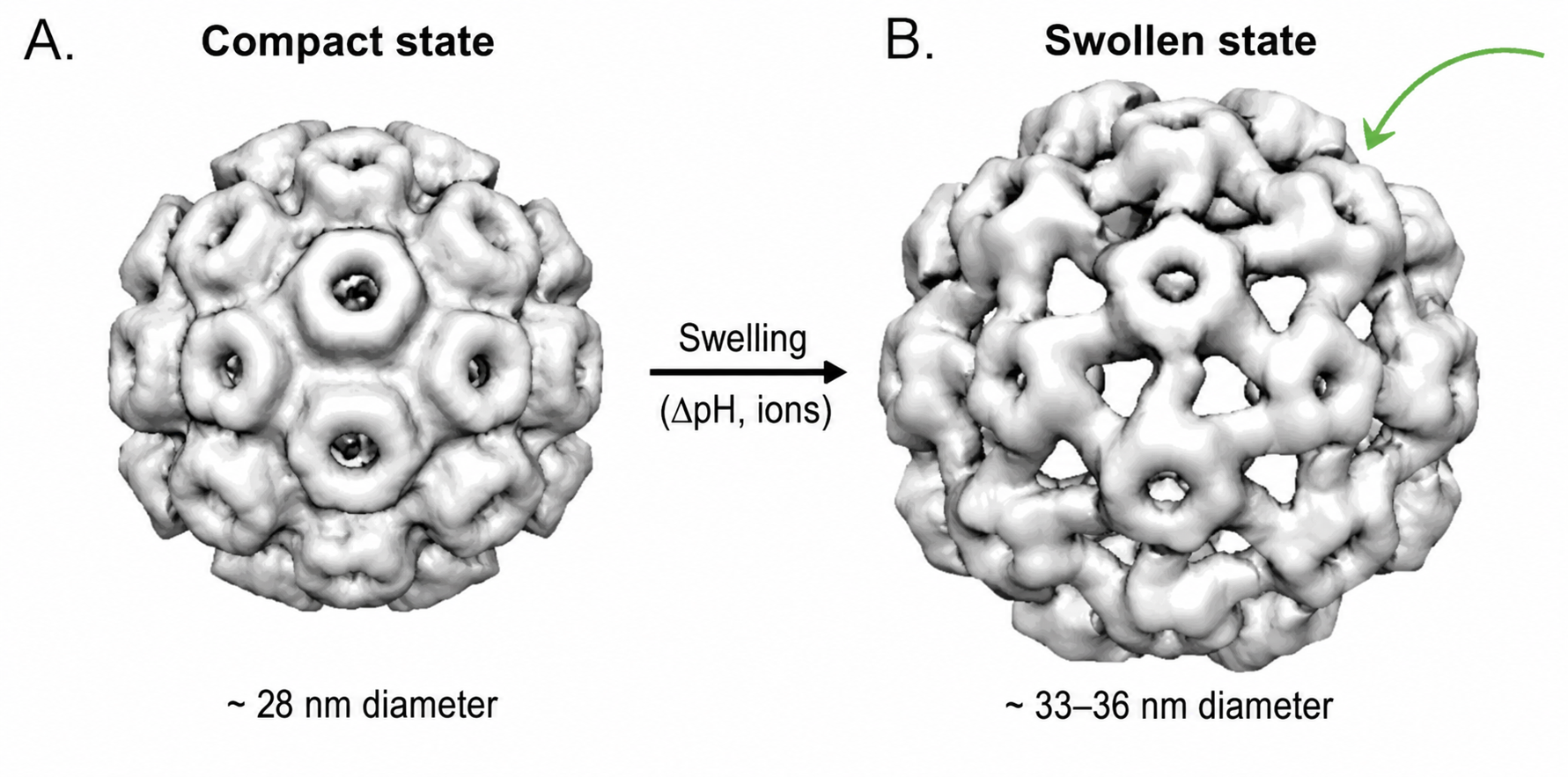}
\caption{A reversible global structural transition is known to exist in several ssRNA icosahedral plant viruses, including BMV \cite{Li2017}. Cycling the pH and the ionic strength allows label access to areas in the swollen structure that are closed in the compact structure (as the green arrow suggests), adapted with permision from ref.~\cite{Speir1995}). }
\label{overv}
\end{figure}

Previously, the all-atom MD simulations of an intact BMV capsid were used determine capsid flexibility and lysine accessibility\cite{AnilSushma2021}. Although  flexibility measurements did not strongly correlate with internal, interfacial, and external designations, molecular accessibility was clearly dependent on lysine position, and in some cases, quasi-equivalence. Specifically, it was found that  most lysines are not sufficiently exposed to accommodate covalent attachment of the fluorophore. Furthermore, simulations predicted low dye accessibility to K86 and 105 and high probability of labeling for K111 while experimental results showed the opposite.

The discrepancy between experimental and computational results can be attributed to difference of pH in labeling reaction and buffer of intact capsid used for simulations. Labeling is facilitated by a reversible swelling transition: the reaction is commonly conducted near pH 6.5-7.5, where BMV becomes more porous, and the capsid structure shows roughly 10\% radial expansion. The largest separation occurs at the quasi-threefold, or trimeric, interfaces, where adjacent capsomers move apart to create pores approximately 2 nm in diameter, giving the expanded shell a more open, truncated-icosahedral appearance while pentamers adopt a flatter orientation relative to the capsid surface as the curvature is redistributed toward the opened trimeric junctions (Appendix Fig.~\ref{overv}). Thus, the newly formed openings expose previously buried protein--protein interfaces and provide access to the capsid lumen; K105 that sits within a depression of the protein surface, underneath K64, in the opening of the fivefold pore becomes more accessible and internal sites such as K86, located beneath the trimeric opening, consequently become accessible to labeling.  Subsequent dialysis into acidic, Mg${^{2+}}$-containing buffer near pH 4.6 restores the compact capsid and reburies interfacial labels.

\subsection{Superradiant dye-conjugated BMV}

Experimentally, heavily labeled BMV particles exhibit the characteristic delayed emission burst and accelerated radiative decay expected for SR \cite{Holmes2022}. These features distinguish SR from incoherent amplified spontaneous emission, which may also occur under room-temperature conditions \cite{Malcuit1987}.

Previous work identified several structural characteristics favorable for SR in OG-BMV. BMV tolerates unusually dense chromophore loading (up to $\sim240$ dyes per particle) without measurable structural distortion \cite{Tsvetkova2019}. At near-saturation labeling, the OG chromophores are found in layers differentiated by radial location and degree of OG accessibility by solvent \cite{AnilSushma2021}.  Reversible swelling of BMV \cite{Incardona1964, Speir1995} allows the labeling of residues located deep within the interior of the virus coat before restoration of the compact state. As a consequence, a subset of labeled sites (K105) has the chromophores strongly confined by the protein environment and largely inaccessible to solvent \cite{krishna2025computational}. This subset of chromophores has the symmetry of the underlying viral architecture, which promotes the indistinguishability of the emitters. In addition, molecular dynamics simulations show that at pH 4.5, the K105 sites exhibit minimal protein flexibility \cite{AnilSushma2021}. SR decreases sharply below saturation labeling, where stochastic labeling breaks this symmetry.

Despite their apparent importance, neither symmetry nor chromophore number alone can explain the observed room-temperature SR. Thus, the swollen capsid retains global symmetry \cite{Speir1995} yet does not exhibit SR. Moreover, comparing the maximum cooperative emission rate, $\Gamma_{SR}^{max}\sim N^2\gamma_0$, with typical collisional dephasing rates ($\sim10^{13}~\mathrm{s^{-1}}$) suggests that emitter numbers substantially larger than those available in BMV would be required for cooperative emission to dominate purely through $N$-scaling. These observations motivated our search for an additional mechanism capable of enhancing collective emission under ambient conditions.

\section{Coupled core--shell oscillator model} \label{SI:CoupledOscillators}

A classical one-dimensional model of two coupled oscillators (Appendix Fig.~\ref{spring}) qualitatively demonstrates how the 'invisible' capsid phonon is reflected by the breathing mode of the AuNP. The first oscillator, representing the AuNP with mass $m_{1}$, is fixed to a wall by a spring with a force constant $k_{1}$, which characterizes its intrinsic oscillation. The second oscillator, representing the capsid with mass $m_{2}$, is coupled to the first via a spring with a force constant $\kappa$ - a characteristic of the core-shell interaction.

We can describe this through the following second-order differential equations. For AuNP:
\begin{equation}
m_1 \ddot{x}_1 = -k_1 x_1 - \kappa (x_1 - x_2) - \gamma_1 \dot{x}_1
\end{equation}
For capsid:
\begin{equation}
m_2 \ddot{x}_2 =  \kappa (x_1 - x_2) - \gamma_2 \dot{x}_2
\end{equation}

where x$_{1,2}$ are the displacements of the two oscillators from their equilibrium. Considering just the breathing modes of AuNP and capsid, x$_{1,2}$ represents the change in their radii. $\gamma_{1,2}$ are the corresponding damping constants. The two masses are known, as well as the effective spring constant for the AuNP ($k_{1} = m_1 \omega_1^2$ from the oscillation frequency of $5.1\text{ cm}^{-1}$ obtained in the experiment (Appendix Table~\ref{edt:core}b). The remaining parameters were used for fitting. Note that the effective spring constants of both the AuNP and the capsid are larger than those obtained from AFM measurements\cite{li2019, Michel2006}. This is not surprising-- in soft biomaterials viscoelastic behavior is characterized by an increase in the mechanical modulus with frequency, often following a power law\cite{SCARCELLI20111539}. Consequently, low-frequency AFM measurements typically yield moduli in the kPa range, whereas GHz Brillouin microscopy yields moduli on the GPa scale\cite{HANDLER2024}

When the pump pulse excites the AuNP, the electronic temperature ($\Delta T_e$) rises sharply~\cite{Hartland2006}. This is followed by rapid electron--electron scattering occurring within $\sim$500~fs~\cite{Link1999}, and electron--phonon scattering in about 1~ps~\cite{Hodak2000}. Subsequently, heat is transferred to the lattice, resulting in a temperature jump ($\Delta T_L$) which leads to expansion and triggers the breathing mode of AuNP. The instantaneous radius change of the AuNP ($\Delta R = x_1$) can be written as a cosine function,
\begin{equation}
    \frac{\Delta R}{R_0}=-\frac{\alpha\Delta T_L}{3}cos(\omega_1 t)+\frac{\alpha\Delta T_L}{3}
\end{equation}
where $R_0$ is the initial radius of AuNP and $\alpha$ is the volumetric expansion coefficient for gold. Calculated from Beer's law and the heat capacity of gold\cite{Hartland2003}, the initial value $\Delta T_e$ was estimated at 46 $K$ (Appendix Table~\ref{edt:core}a). Since $\Delta T_e\approx1000 K$ will only lead to $\Delta T_L\approx20 K$\cite{Hartland2006}, $\Delta T_L$ here is chosen as 2 $K$. 

The initial conditions to solve the system of differential equations are set as follows. The clocking of this model begins at a pump--probe delay of $t_0 \approx 3.25\,\text{ps}$; at the midpoint of the first period, where the AuNP expansion reaches its maximum with zero velocity. The virus capsid is set to have a phase lag of $\pi/2$ with the maximum velocity. 

Thus, for the AuNP, the initial conditions are: 
\begin{align*}
   x_1(t_0)&=\frac{2\alpha\Delta T_L}{3} R_0\\
    \dot{x_1}(t_0)&= 0
\end{align*}

For capsid, taking into account momentum conservation:   
\begin{align*}
    x_2(t_0)&=(m_1/m_2)\frac{\alpha\Delta T_L}{3} R_0=13.3\frac{\alpha\Delta T_L}{3} R_0\\
    \dot{x_2}(t_0)&= (m_1/m_2)\frac{\alpha\Delta T_L}{3}\omega_1R_0=13.3\frac{\alpha\Delta T_L}{3}\omega_1R_0
\end{align*}

The radius change $\Delta R(t)$ was numerically solved using ode45s in MATLAB.

\begin{table*}[!ht]
\caption{\textbf{Gold-core heating estimates and oscillator parameters.}}
\label{edt:core}
\begingroup\setstretch{1.15}\small
\noindent \textbf{a} \textbf{Electronic temperature estimation.}\par\smallskip
\begin{center}\begin{adjustbox}{max width=\textwidth}\begin{tabular}{l l}
        \hline
        \textbf{Description} & \textbf{Value} \\
        \hline
        Energy per pump pulse & $9.6\times 10^{-6}\,\mathrm{mJ}$ \\
        Energy density per pump& $1.1\times 10^{-9}\,\mathrm{mJ}\,\mu\mathrm{m}^{-2}$ \\
        Absorption coefficient at 515\,nm (20\,nm AuNP) & $1.06\times 10^{9}\,\mathrm{M^{-1}\,cm^{-1}}$ \\
        Number of Au atoms per AuNP & $2.48\times 10^{5}$ \\
        Absorption coefficient at 515\,nm (Au atom) & $4.28\times 10^{3}\,\mathrm{M^{-1}\,cm^{-1}}$ \\
        Molar absorbed energy per pump & $1.08\times 10^{3}\,\mathrm{J\,mol^{-1}}$ \\
        Au atom electronic temperature increase& $46\,\mathrm{K}$ \\
        \hline
    \end{tabular}\end{adjustbox}\end{center}

\noindent \textbf{b} \textbf{Coefficients used in the coupled oscillator model.} 
    Parameters of the AuNP--capsid model.\par\smallskip
\begin{center}\begin{adjustbox}{max width=\textwidth}\begin{tabular}{l l l}
        \hline
        \textbf{Symbol} & \textbf{Description} & \textbf{Value} \\
        \hline
        $m_{1}$       & AuNP mass                & $8 \times 10^{-20}\,\mathrm{kg}$ \\
        $m_{2}$       & Capsid mass              & $6 \times 10^{-21}\,\mathrm{kg}$ \\
        $k_{1}$       & AuNP spring constant     & $7.4 \times 10^{4}\,\mathrm{N/m}$ \\
        $\kappa$      & Coupling constant        & $3.6 \times 10^{2}\,\mathrm{N/m}$ \\
        $\gamma_{1}$  & AuNP damping coefficient & $1.2 \times 10^{-8}\,\mathrm{kg/s}$ \\
        $\gamma_{2}$  & Capsid damping coefficient & $3 \times 10^{-10}\,\mathrm{kg/s}$ \\
        \hline
    \end{tabular}\end{adjustbox}\end{center}
\endgroup
\end{table*}

\section{Continuum estimates of acoustic modes}\label{SI:Identification}

The two experimental approaches of this study reveal low-frequency oscillations near \(1~\mathrm{cm^{-1}}\): \(\approx 1.3~\mathrm{cm^{-1}}\) in AuVLPs and \(\approx 1.0~\mathrm{cm^{-1}}\) in dye-saturated BOG309 particles. Given the limited spectral resolution of the TA measurements, these frequencies should be regarded as belonging to the same low-frequency acoustic range rather than as sharply resolved mode frequencies. 

In the AuVLP experiment, the AuNP breathing mode provides an approximately isotropic radial drive. After optical excitation of the Au core, impulsive thermoelastic expansion launches the \(l=0\) breathing mode of the nanoparticle, which applies a radial stress to the surrounding capsid. Therefore, one may expect the capsid response to acquire an initially predominantly \(l=0\) character. However, protein assemblies cannot be treated as purely harmonic springs beyond small amplitudes \cite{Miyashita2003}. Moreover, one expects a wavepacket, not a single mode, because of the impulsive nature of the excitation.  

In BOGs,  since the dye transition dipoles are presumably organized according to the capsid structure, the excited-state structural relaxation is expected to exert a local force on the protein scaffold which can generate an anisotropic stress pattern. Thus, dye excitation can in principle drive not only a fully symmetric  breathing-like response, but also lower-symmetry acoustic modes. Some of these modes will be more damped than the others. To study their frequencies and relaxation times, we have approached the problem in two ways: one based on a continuum treatment, and the other using a discrete coarse-grained model. Although real capsids have discrete icosahedral symmetry, the lowest acoustic modes are often well captured by continuum elastic models and can subsequently be decomposed into icosahedral representations \cite{YANG2009,Dykeman2010}. Therefore the two approaches should provide consistent, although not overlapping, results.

In the continuum elastic-sphere model, the mode frequencies are determined primarily by the particle size, longitudinal sound velocity \(v_L\), and Poisson ratio \(\sigma\) \cite{Ford2003,Ford2003corr,Talati2006}.  Following prior work on virus and protein nanoparticles, we used \(v_L = 1817~\mathrm{m\,s^{-1}}\), estimated from lysozyme and virus crystals, and \(\sigma = 0.33\), typical of soft condensed materials \cite{TACHIBANA2000,Stephanidis2007Elastic,YANG2009,saviot2007virus}.

We modeled BOG309 as a homogeneous 28 nm protein sphere and AuVLP as a core--shell particle consisting of a 20 nm Au core surrounded by a 10 nm-thick capsid shell. Including damping by water, we estimated the frequencies and lifetimes of the lowest acoustic modes for both particles, as summarized in Appendix Table~\ref{edt:models}b \cite{Saviot2007Damping,saviot_cfm_calculator}. The AuVLP modulation at \(\approx 1.3~\mathrm{cm^{-1}}\) is consistent with the calculated \(l=0\) core--shell mode, supporting its assignment to a capsid-associated breathing-like response mechanically coupled to the AuNP.

 The fully symmetric breathing mode (l=0) frequency corroborates well with our experimental results. However, the (l=2) quadrupolar mode has a predicted lifetime that is more consistent with our experimental observations. 

We estimate the lowest acoustic phonon modes of AuVLP and BOG309 by modeling BOG309 as a homogeneous sphere and AuVLP as a core--shell particle with a 20~nm gold core and a 10~nm-thick capsid \cite{Saviot2007Damping,saviot_cfm_calculator}. Material constants are listed in Appendix Table~\ref{edt:models}c.

\section{Model of vibrationally assisted emission}\label{SI:Qmodel}
\subsection{Capsid vibrational spectrum}\label{SI:spectrum}
In Appendix Fig.~\ref{capsid_normalmodes}C, we show the density of the vibrational modes from the elastic-network model, which forms two peaks centered around 1.1 $\text{cm}^{-1}$ and 2.5 $\text{cm}^{-1}$. We divided the capsid vibrations into under-damped modes, which have relaxation times that are commensurate or longer than their periods, and over-damped modes, which have relaxation times that are much shorter than their periods. Both categories of vibrational modes coexist throughout the frequency range, except a gap from 1.4 $\text{cm}^{-1}$ to 2.3 $\text{cm}^{-1}$, which is exclusively populated by a low density of isolated under-damped modes. In Panel D, we present the lowest frequency vibrational modes from each degeneracy group $n$, demonstrating instances of an over-damped mode, such as the lowest frequency breathing mode (first panel), and three under-damped modes (the rest three panels) that form the low-frequency tail of the vibrational mode density. In Section~\ref{VibAssSR}, we incorporate aspects of the over-damped and under-damped vibrational dynamics of the capsid in the model of vibrationally assisted superradiance. Specifically, we model the stochastic fluctuations originating from the over-damped modes via a diffusive, Orstein-Uhlenbeck process, and we explicitly incorporate the coupling of chromophores to an effective under-damped mode. Such under-damped modes are seen to populate nearly uniformly the capsid vibrational spectrum and to occupy low- and intermediate-frequency windows that are depleted from over-damped modes.      

\subsection{Vibrationally assisted superradiance} \label{VibAssSR}
We describe the chromophores as two-level systems consisting of a ground $|0_i\rangle$ and an excited state $|1_i\rangle$ and express the model of superradiance in terms of raising (absorption) $\sigma_{+,i}=|1_i \rangle \langle 0_i |$ and lowering (emission) $\sigma_{-,i}=|0_i \rangle \langle 1_i |$ operators of chromophore $i$. The product $\sigma_{+,i}\sigma_{-,i} = |1_i \rangle \langle 1_i |$ is the projection operator on the excited state of the chromophore. We write the Hamiltonian of the chromophore system interacting with the capsid vibrational motion, $H_{CV}$, as
\begin{equation}\label{superrH}
\begin{split}
    H_{CV} &= \sum_i \hbar \Omega_{E,i} \sigma_{+,i}\sigma_{-,i} + \hbar \Omega_{EE} \sum_{i \neq j} \sigma_{+,i}\sigma_{-,j} + \\ &+ \hbar \Omega_{EPh} \sum_i \sigma_{+,i}\sigma_{-,i} x  + \hbar \Omega_{EEPh} \sum_{i \neq j} \sigma_{+,i}\sigma_{-,j} x + \\ & +  \frac{\hbar \Omega_{Ph}}{2} \left( p^2 + x^2 \right),
\end{split} 
\end{equation}
where the sums over $i$ and $j$ go over all chromophores in the system, and the Hamiltonian is expressed in the rotating frame of the optical transition. 

The characterization of the chromophore intra-molecular geometric changes upon excitation showed that the electronic transition of the dye couples strongly to the intra-molecular rotation. Moreover, this rotation will interact with the capsid motion via fluctuations of the coat protein separations that are driven by both over-damped and under-damped vibrational modes. To account for the interaction of the chromophores with the over-damped vibrational motion, the first term in Eq.~\ref{superrH} includes the chromophore-specific detuning, $\hbar \Omega_{E,i}$, of the transition energy from resonance\cite{kubo1969stochastic}. We assume that $\Omega_{E,i}$ are independent Ornstein-Uhlenbeck (OU) stochastic processes\cite{uhlenbeck1930theory}, governed by the stochastic differential equations\cite{kubo1963stochastic,gillespie1996exact}
\begin{equation}\label{oueqn}
\dot{\Omega}_{E,i} = -\eta \Omega_{E,i} + \sigma dW_i
\end{equation} 
with relaxation rate $\eta$ and variance $\sigma^2 = 2\eta\Delta^2$, which characterize the over-damped stochastic fluctuations of the viral capsid, and $dW_i$ are Wiener processes with zero mean and unit variance. We derive $\eta$ from simulations of the coat protein-protein distance autocorrelation function with the Langevin elastic-network model for capsid vibrational dynamics, see Appendix Fig.~\ref{capsid_normalmodes}B, and derive the OU broadening, $\Delta$, according to the experimentally measured superradiance broadening between 510nm and 520nm.

The second term in Eq.~\ref{superrH} represents the exciton-exciton interaction with  strength $\hbar \Omega_{EE}$, which we estimated from the strength of the dipole-dipole interaction of two parallel chromophores with $10$~D transition dipole moments at $2$~nm distance. The next two terms describe the coupling of the excitons to a collective, underdamped vibration with dimensionless displacement $x$. $\hbar \Omega_{EPh}$ characterizes the coupling strength of the exciton transition to the vibrational motion, which we took to be proportional to the experimentally observed superradiance broadening. $\hbar \Omega_{EEPh}$ characterizes the coupling strength of the exciton interaction to the vibrational motion, which we estimated from the magnitude of the derivative of the dipole-dipole interaction of two parallel nearest neighbor chromophores, multiplied by the root-mean-square deviation, 0.035 nm, of the distance of neighboring coat proteins. The last term in Eq.~\ref{superrH} is the Hamiltonian of the underdamped vibrational motion with characteristic frequency $\Omega_{Ph}$ and dimensionless momentum $p$. We adjusted the value of $\Omega_{Ph}$ to reproduce the period of the oscillations of the superradiance interference pattern. The obtained value of $\Omega_{Ph}$ falls well within the range of under-damped vibrations of the capsid, computed with the Langevin elastic network model. The parameters of the model of superradiance are collected in Appendix Table~\ref{edt:models}a.

We simulated the process of superradiance from the multi-chromophore system using an open system approach with a quantum master equation given by\cite{Gross1982a,temnov2005superradiance}
\begin{equation}
    \dot{\rho} = -\frac{i}{\hbar} \left[ H_{CV}(t), \rho \right] + \Gamma \left( S_{-} \rho S_{+} - \frac{1}{2} \left\{ S_{+} S_{-}, \rho \right\} \right),
\end{equation}
where $\rho$ is the chromophore-system density matrix; $H_{CV}$ - the chromophore-vibration Hamiltonian, which explicitly depends on time via the stochastic OU processes, $\Omega_{E,i}$, and the collective, underdamped vibrational coordinate, $x$; $S_{-} = \sum_i \sigma_{-,i}$ - the collective lowering (emission) operator; $S_{+} = S^{\dagger}_{-}$ - the collective raising (absorption) operator; $\Gamma$ - the single chromophore emission rate, which we adjusted to obtain a superradiant peak at the experimentally observed 20ps delay with a 10 chromophore system. $[\cdot, \cdot]$ stands for a commutator and $\{\cdot, \cdot\}$ - for an anti-commutator.

We simulated the underdamped vibrational motion in the self-consistent, mean-field (Ehrenfest) approximation\cite{tully1991nonadiabatic}, where the force exerted on the harmonic oscillator derives from the expectation value of the chromophore system Hamiltonian, and the vibrational motion feeds back into the quantum evolution via the time dependence of the vibrational coordinate. The classical Hamilton equations of motion for the underdamped vibration together with a damping term to account for its finite lifetime are
\begin{equation}
\begin{split}
\dot{p} &= - \eta_{Ph} p - \frac{1}{\hbar} \frac{\partial \langle H_{CV} \rangle }{ \partial x} = ,\\
&= - \eta_{Ph} p - \Omega_{Ph} x - \Omega_{EPh} \langle \sum_i \sigma_{+,i}\sigma_{-,i} \rangle - \Omega_{EEPh} \langle \sum_{i \neq j} \sigma_{+,i}\sigma_{-,j} \rangle ,\\
\dot{x} &= \Omega_{Ph} p,
\end{split}
\end{equation}
where $\langle \cdot \rangle = \text{Tr} [\rho \cdot]$ is the trace of quantum operators with the time-dependent chromophore-system density matrix, and $\eta_{Ph}$ is the mode relaxation rate, which we obtained from the vibrational analysis of the capsid.

We integrated the superradiance quantum master equation in the interaction representation, which allowed us to use a longer integration time step. The construction of the interaction representation required the diagonalization of the chromophore-vibration Hamiltonian in the product chromophore basis at each time step, which limited the number of chromophores we can treat to 10. We initiated the quantum evolution from a product state, where all chromophores are in an excited state and are at resonance. We similarly initiated the underdamped vibration at the equilibrium position $x=0$ and $p=0$. We solved simultaneously the superradiance quantum master equation and the classical Hamilton equations, and we computed the expectation value $\text{Tr}[\rho S_{+} S_{-}]$ as a function of time, which is proportional to the intensity of emission from the chromophore system. The results are shown in the main text, Figure~\ref{mechanism_superradiance}.

\subsection{Synchronization in a system of two chromophores}
\label{SI:TwinSync}
To obtain further insight in the mechanism it is useful to consider the emission from a system of just two interacting chromophores. The Hamiltonian given by:
\begin{equation}\label{2chromH}
H = \hbar \Omega_{E,1} \sigma_{+,1}\sigma_{-,1} + \hbar \Omega_{E,2} \sigma_{+,2}\sigma_{-,2} + \hbar \Omega_{EE} (\sigma_{+,1}\sigma_{-,2} + \sigma_{+,2}\sigma_{-,1}), 
\end{equation}
where $\Omega_{E,1}$ and $\Omega_{E,2}$ are the detuning of the chromophore energies from resonance, and $\Omega_{EE}$ is the strength of the inter-chromophore interaction. The four states of the two chromophores fall into a triplet of emissive, symmetric states
\begin{equation}
\begin{split}
&| 1_1 1_2 \rangle, \\
&| + \rangle = \frac{1}{\sqrt{2}} \left( | 1_1 0_2 \rangle + | 0_1 1_2 \rangle \right), \\
&| 0_1 0_2 \rangle,
\end{split}
\end{equation}
which includes the superradiant state $| + \rangle$, and a singlet of a dark, subradiant state
\begin{equation}
\begin{split}
| - \rangle = \frac{1}{\sqrt{2}} \left( | 1_1 0_2 \rangle - | 0_1 1_2 \rangle \right).
\end{split}
\end{equation}
Here, the states are labeled by the presence $|1\rangle$ or absence $|0\rangle$ of an exciton on a chromophore. In an optical experiment, we initially excite the two chromophores in the product state $| 1_1 1_2 \rangle$, which then cascades down the states of the triplet upon consecutive one-photon emissions. Superradiance emerges from this cascade by the formation of the superradiant state $|+\rangle$ due to the indistinguishable emission of photons. Interactions that admix the subradiant $|-\rangle$ and the superradiant $|+\rangle$ states, however, destroy the constructive interference of excitons and suppress superradiance. For example, in the case of an equal superposition of the two  states, we obtain a product chromophore state, which emits incoherently 
\begin{equation}
  | 1_1 0_2 \rangle = \frac{1}{\sqrt{2}} \left( | + \rangle + | - \rangle \right).  
\end{equation}
To infer how the interactions in the two chromophore system influence this superposition, we form the sub-matrix of the two-chromophore Hamiltonian in Eq.~\ref{2chromH} between the super- and subradiant states
\begin{equation}\label{2chromophoreH}
\begin{bmatrix}
\hbar \Omega^+_E + \hbar \Omega_{EE} & -\hbar \Omega^-_E \\
-\hbar \Omega^-_E & \hbar \Omega^+_E - \hbar \Omega_{EE}
\end{bmatrix},
\end{equation}
where $\Omega^+_E = \frac{\Omega_{E,1} + \Omega_{E,2}}{2}$ and $\Omega^-_E = \frac{\Omega_{E,2} - \Omega_{E,1}}{2}$. The off-diagonal terms in Eq.~\ref{2chromophoreH}, which contain the difference of the chromophore detuning, couple the two states, resulting in the suppression of superradiance and incoherent emission. Their decoherence effect is counteracted by the exciton interaction, which opens an energetic gap between the two states, resulting in diminished coupling between them and restoration of superradiance. Synchronous shifts in the chromophore detuning do not influence the superposition of the two states and therefore have no effect on superradiance.

Extending this analysis to the time-dependent two-chromophore Hamiltonian reveals that the uncorrelated, stochastic fluctuations in chromophore detuning in Eq.~\ref{superrH}, arising from the over-damped dynamics of the capsid, produce incoherent emission through coupling with the subradiant states. The analysis further clarifies the mechanism of vibrationally assisted superradiance mediated by the under-damped capsid motion. Excitation of the chromophores displaces the collective vibrational coordinate from equilibrium - owing to the strong exciton-vibrational coupling represented by the third term in Eq.~\ref{superrH} - initiating an under-damped oscillation. This coherent vibrational motion, in turn, collectively modulates the exciton-exciton interactions (the fourth term in Eq.~\ref{superrH}), opening an energetic gap between the superradiant and subradiant manifolds. This gap suppresses the coupling between the two manifolds and restores superradiance. Therefore, the essential new ingredient in the model is the self-consistent interaction between the chromophore system and the vibrational motion: the chromophore system drives the vibrational motion upon excitation, and the resulting vibrational dynamics feeds back by modulating the inter-chromophore interactions. 

This dynamic, self-consistent exciton-vibrational coupling produces the two defining spectroscopic features of the observed superradiance: (i) the oscillation of the under-damped vibrational coordinate correlates with the collective oscillations in chromophore detuning, producing the observed oscillations in the transient absorption experiment; (ii) the same vibrational oscillation drives oscillations in the energetic gap between the superradiant and subradiant states, producing the observed oscillatory pattern of the superradiant emission - the emission is strongest near the turning points of the vibrational motion, where the energetic gap is largest, and it is weakest near the equilibrium position, where the gap vanishes.

\bibliography{BDsRefs}

\end{document}